\documentclass[10pt,journal,compsoc]{IEEEtran}

\usepackage{amsmath} %
\usepackage{amsthm}
\theoremstyle{definition}

\usepackage{listings}  
\usepackage{booktabs} 
\usepackage{enumitem}
\usepackage{multirow}
\usepackage{makecell}
\usepackage[frozencache]{minted}
\usepackage{algorithm}
\usepackage{algorithmicx}
\usepackage[noend]{algpseudocode}
\usepackage{cite}
\algrenewcommand{\algorithmiccomment}[1]{{\hfill \textcolor{gray}{$\triangleright$ #1}}}

\setminted{xleftmargin=\parindent}

\newcommand{\toolname}{\testpilot}

\usepackage{authorcomments}
\usepackage{subcaption}
\usepackage[export]{adjustbox}

\usepackage{hyperref}

\usepackage[scaled=0.875]{inconsolata}

\newcommand{\code}[1]{\text{\lstinline[basicstyle=\ttfamily\small, language=JavaScriptColor]~#1~}}

\newcommand{\exploreAPI}[1]{\mbox{\it exploreAPI(#1)}\xspace}

\newcommand{\sig}{\mbox{\it sig}\xspace}

\newcommand{\snippets}{\mbox{\it snippets}\xspace}

\newcommand{\accessPath}{\mbox{\it accessPath}\xspace}

\newcommand{\Pair}[2]{\mbox{$\langle$}#1,#2\mbox{$\rangle$}\xspace}
\newcommand{\seen}{\mbox{\it seen}\xspace}
\newcommand{\apis}{\mbox{\it apis}\xspace}
\newcommand{\api}{\mbox{\it api}\xspace}
\newcommand{\prop}{\mbox{\it prop}\xspace}
\newcommand{\props}{\mbox{\it props}\xspace}
\newcommand{\obj}{\mbox{\it obj}\xspace}
\newcommand{\pkgName}{\mbox{\it pkgName}\xspace}
\newcommand{\modObj}{\mbox{\it modObj}\xspace}

\newcommand{\explore}[3]{\mbox{$\mbox{\it explore}(#1,#2,#3)$}\xspace}
\newcommand{\extend}{\mbox{\it extend}\xspace}
\newcommand{\component}{\mbox{\it component}\xspace}

\definecolor{javared}{rgb}{0.6,0,0} %
\definecolor{javagreen}{rgb}{0.25,0.5,0.35} %
\definecolor{javapurple}{rgb}{0.5,0,0.35} %
\definecolor{javadocblue}{rgb}{0.25,0.35,0.75} %

\lstdefinelanguage{JavaScriptColor}{
  keywords={await, async, break, case, catch, const, continue, debugger, default, delete, do, else,
    export, finally, for, function, if, import, in, instanceof, let, of, new, null, return, require, switch, this,
    throw, try, typeof, var, void, while, with, super,
    class, interface, implements, public, private, constructor
    },
  morecomment=[l]{//},
  morecomment=[s]{/*}{*/},
  morestring=[b]',
  morestring=[b]",
  keywordstyle=\color{javapurple}\bfseries,
  identifierstyle=\color{black},
  commentstyle=\color{javagreen}\ttfamily,
  numberstyle=\color{javared}\ttfamily,
  stringstyle=\color{javared}\ttfamily,
  sensitive=false,
  numbers=left,
  stepnumber=1,
  escapeinside={/*\#}{\#*/},
}
\newcommand{\testpilot}{\textsc{TestPilot}\xspace}

\newcommand{\FunctionBodyIncluder}{\textit{FnBodyIncluder}\xspace}
\newcommand{\DocCommentIncluder}{\textit{DocCommentIncluder}\xspace}
\newcommand{\RetryWithError}{\textit{RetryWithError}\xspace}
\newcommand{\SnippetIncluder}{\textit{SnippetIncluder}\xspace}

\usepackage{tcolorbox}

\newcommand{\cushman}{\textit{code-cushman-002}\xspace}
\newcommand{\gptturbo}{\textit{gpt3.5-turbo}\xspace}
\newcommand{\starcoder}{\textit{StarCoder}\xspace}

\newcommand{\prompts}{\mbox{\it prompts}\xspace}
\newcommand{\promptsWithFnBody}{\mbox{\it promptsWithFnBody}\xspace}
\newcommand{\promptsWithExampleSnippets}{\mbox{\it promptsWithExamples}\xspace}
\newcommand{\promptsWithDocComments}{\mbox{\it promptsWithDocComments}\xspace}

\newcommand{\prompt}{\mbox{\it prompt}\xspace}
\newcommand{\newPrompt}{\mbox{\it prompt$'$}\xspace}

\newcommand{\body}{\mbox{\it body}\xspace}
\newcommand{\docComment}{\mbox{\it docComment}\xspace}
\newcommand{\LLM}{\mbox{\it LLM}\xspace}
\newcommand{\completion}{\mbox{\it completion}\xspace}
\newcommand{\completions}{\mbox{\it completions}\xspace}
\newcommand{\test}{\mbox{\it test}\xspace}
\newcommand{\tests}{\mbox{\it tests}\xspace}
\newcommand{\crash}{\mbox{\it crash}\xspace}
\newcommand{\nonTermination}{\mbox{\it nonTermination}\xspace}

\newcommand{\result}{\mbox{\it result}\xspace}
\newcommand{\ok}{\mbox{\it ok}\xspace}
\newcommand{\resultStatus}{\mbox{\it result}.\mbox{\it status}\xspace}
\newcommand{\resultErrorMessage}{\mbox{\it result}.\mbox{\it errorMessage}\xspace}

\newcommand{\assertionFailure}{\mbox{\it assertionFailure}\xspace}
\newcommand{\createBasePrompt}[1]{\mbox{\it createBasePrompt}(#1)\xspace}
\newcommand{\findExamplesForSig}[2]{\mbox{\it findExamples}(#1,~#2)\xspace}
\newcommand{\findFunctionBodyForSig}[2]{\mbox{\it findFnBody}(#1,~#2)\xspace}
\newcommand{\findDocCommentsForSig}[2]{\mbox{\it findDocComments}(#1,~#2)\xspace}

\newcommand{\refine}[2]{\mbox{\it refine}(#1,\:#2)\xspace}
\newcommand{\refineFromError}[2]{\mbox{\it refineFromError}(#1,\:#2)\xspace}
\newcommand{\getCompletions}[2]{\mbox{\it getCompletions}(#1,\:#2)\xspace}
\newcommand{\concatenate}[2]{\mbox{\it concatenate}(#1,\:#2)\xspace}
\newcommand{\executeTest}[1]{\mbox{\it executeTest}(#1)\xspace}
\newcommand{\removeComments}[1]{\mbox{\it removeComments}(#1)\xspace}
\newcommand{\fixMinorSyntaxIssues}[1]{\mbox{\it fixMinorSyntaxIssues}(#1)\xspace}

\newcommand{\Set}[1]{\mbox{$\{\:$}#1\mbox{$\:\}$}\xspace}

\newcommand{\cushmanmedianstmtCoverage}{68.2\%\xspace}

\newcommand{\cushmanmedianbranchCoverage}{51.2\%\xspace}

\newcommand{\cushmanmedianTotalTime}{4m 53s\xspace}

\newcommand{\cushmanmedianAvgTimePerMethod}{11s\xspace}

\newcommand{\gptturbonumPackages}{25\xspace}
\newcommand{\gptturbonumPackagesWithDocComments}{8\xspace}

\newcommand{\gptturbonumTotalApiFunctions}{1,684\xspace}
\newcommand{\gptturbominstmtCoverage}{33.9\%\xspace}
\newcommand{\gptturbomaxstmtCoverage}{93.1\%\xspace}
\newcommand{\gptturbomedianstmtCoverage}{70.2\%\xspace}

\newcommand{\gptturbominGitLabstmtCoverage}{51.4\%\xspace}
\newcommand{\gptturbomaxGitLabstmtCoverage}{78.3\%\xspace}

\newcommand{\gptturbominbranchCoverage}{16.5\%\xspace}
\newcommand{\gptturbomaxbranchCoverage}{71.3\%\xspace}
\newcommand{\gptturbomedianbranchCoverage}{52.8\%\xspace}

\newcommand{\gptturbomediannonTrivialCoverage}{61.6\%\xspace}

\newcommand{\gptturbonumNessieZeroBranchCov}{3\xspace}
\newcommand{\gptturboNessieZeroBranchCovProjs}{dirty, geo-point, core\xspace}
\newcommand{\gptturbominNessiestmtCoverage}{4.7\%\xspace}
\newcommand{\gptturbomaxNessiestmtCoverage}{96.0\%\xspace}
\newcommand{\gptturbomedianNessiestmtCoverage}{51.3\%\xspace}

\newcommand{\gptturbonumProjTestPilotHigherThanNessiestmtCoverage}{17\xspace}
\newcommand{\gptturboprojsTestPilotHigherThanNessiestmtCoverage}{glob, fs-extra, bluebird, q, rsvp, memfs, js-sdsl, quill-delta, complex.js, pull-stream, simple-statistics, plural, dirty, geo-point, image-downloader, core, omnitool\xspace}
\newcommand{\gptturbomintpVsNessieHigherstmtCoverageDiff}{3.6\%\xspace}
\newcommand{\gptturbomaxtpVsNessieHigherstmtCoverageDiff}{74.5\%\xspace}
\newcommand{\gptturbomediantpVsNessieHigherstmtCoverageDiff}{30.0\%\xspace}

\newcommand{\gptturbonumProjTestPilotLowerThanNessiestmtCoverage}{7\xspace}
\newcommand{\gptturboprojsTestPilotLowerThanNessiestmtCoverage}{graceful-fs, jsonfile, node-dir, zip-a-folder, countries-and-timezones, crawler-url-parser, gitlab-js\xspace}
\newcommand{\gptturbomintpVsNessieLowerstmtCoverageDiff}{0.5\%\xspace}
\newcommand{\gptturbomaxtpVsNessieLowerstmtCoverageDiff}{53.2\%\xspace}
\newcommand{\gptturbomediantpVsNessieLowerstmtCoverageDiff}{3.6\%\xspace}

\newcommand{\gptturbomedianNessiebranchCoverage}{25.6\%\xspace}

\newcommand{\gptturbominPercentPassing}{9.9\%\xspace}
\newcommand{\gptturbomaxPercentPassing}{80.0\%\xspace}
\newcommand{\gptturbomedianPercentPassing}{48.0\%\xspace}

\newcommand{\gptturbomaxPercentUniquelyCoveringTests}{100.0\%\xspace}
\newcommand{\gptturbomedianPercentUniquelyCoveringTests}{10.5\%\xspace}

\newcommand{\gptturboremainingNonUniqueTests}{89.5\%\xspace}

\newcommand{\gptturbomaxPercentNonTrivialTests}{94.6\%\xspace}
\newcommand{\gptturbomedianPercentNonTrivialTests}{61.4\%\xspace}

\newcommand{\gptturbosecondMinPercentNonTrivialTests}{9.1\%\xspace}

\newcommand{\gptturbomedianPercentNonTrivialTestsPassing}{43.7\%\xspace}

\newcommand{\gptturbominDiffAllVsNonTrivial}{0.0\%\xspace}
\newcommand{\gptturbomaxDiffAllVsNonTrivial}{84.0\%\xspace}
\newcommand{\gptturbomedianDiffAllVsNonTrivial}{7.5\%\xspace}

\newcommand{\gptturbonumPkgsWithZeroNonTrivialCoverage}{4\xspace}
\newcommand{\gptturbopkgsWithZeroNonTrivialCoverage}{jsonfile, node-dir, zip-a-folder, image-downloader\xspace}

\newcommand{\gptturbomedianTotalTime}{6m 55s\xspace}

\newcommand{\gptturbomedianAvgTimePerMethod}{15s\xspace}

\newcommand{\gptturbomedianPercentTimeoutErrors}{22.7\%\xspace}

\newcommand{\gptturbomedianPercentAssertionErrors}{19.2\%\xspace}

\newcommand{\gptturbomedianPercentCorrectnessErrors}{20.0\%\xspace}

\newcommand{\gptturbomindiffstmtCoverageWithLoading}{19.1\%\xspace}
\newcommand{\gptturbomaxdiffstmtCoverageWithLoading}{88.2\%\xspace}
\newcommand{\gptturbomediandiffstmtCoverageWithLoading}{53.7\%\xspace}

\newcommand{\gptturbomindiffbranchCoverageWithLoading}{15.9\%\xspace}
\newcommand{\gptturbomaxdiffbranchCoverageWithLoading}{71.3\%\xspace}
\newcommand{\gptturbomediandiffbranchCoverageWithLoading}{50.0\%\xspace}

\newcommand{\gptturbonumPkgsWithUselessRefiner}{4\xspace}

\newcommand{\gptturbosimLEQthirty}{6.2\%\xspace}
\newcommand{\gptturbosimLEQforty}{60.0\%\xspace}
\newcommand{\gptturbosimLEQfifty}{92.8\%\xspace}
\newcommand{\gptturbosimLEQsixty}{99.6\%\xspace}
\newcommand{\gptturbosimLEQseventy}{100.0\%\xspace}

\newcommand{\gptturbosimGTfifty}{7.2\%\xspace}

\newcommand{\gptturbominMedianFnCovPerProj}{0.0\%\xspace}
\newcommand{\gptturbomaxMedianFnCovPerProj}{100.0\%\xspace}
\newcommand{\gptturbomedianMedianFnCovPerProj}{77.1\%\xspace}
\newcommand{\gptturboAssertionErrorFixedWithRetry}{11.1\%\xspace}

\newcommand{\gptturboTimeoutErrorFixedWithRetry}{15.4\%\xspace}

\newcommand{\gptturboallErrorsFixedWithRetry}{15.6\%\xspace}

\newcommand{\starcodermedianstmtCoverage}{54.0\%\xspace}

\newcommand{\starcodermedianbranchCoverage}{37.5\%\xspace}

\newcommand{\starcodermedianNessiestmtCoverage}{51.3\%\xspace}

\newcommand{\starcodermedianNessiebranchCoverage}{25.6\%\xspace}

\newcommand{\starcodermedianTotalTime}{10m 48s\xspace}

\newcommand{\starcodermedianAvgTimePerMethod}{24s\xspace}

\begin{document}

\title{An Empirical Evaluation of Using Large Language Models 
for Automated Unit Test Generation}

\author{
    Max Sch\"afer, 
    Sarah Nadi, 
    Aryaz Eghbali,
    Frank Tip%
\IEEEcompsocitemizethanks{
\IEEEcompsocthanksitem M. Sch\"afer is with GitHub, UK\protect\\
E-mail: max-schaefer@github.com
\IEEEcompsocthanksitem S. Nadi is with the University of Alberta, Canada\protect\\
E-mail: nadi@ualberta.ca
\IEEEcompsocthanksitem A. Eghbali is with the University of Stuttgart, Germany\protect\\
E-mail: aryaz.egh@gmail.com
\IEEEcompsocthanksitem F. Tip is with  Northeastern University, USA\protect\\
E-mail: f.tip@northeastern.edu
}}

\IEEEtitleabstractindextext{
\begin{abstract}
Unit tests play a key role in ensuring the correctness of software.
However, manually creating unit tests is a laborious task, motivating the need for automation.
Large Language Models (LLMs) have recently been applied to various aspects of software development,
including their suggested use for automated generation of unit tests, but while requiring additional training or few-shot learning on examples of existing tests.
This paper presents a large-scale empirical evaluation on the effectiveness of LLMs for
automated unit test generation without requiring additional training or manual effort.
Concretely, we consider an approach where the LLM is provided with prompts that include the signature and implementation of a function under test, along with usage examples extracted from documentation.
Furthermore, if a generated test fails, our approach attempts to generate a new test that fixes the problem
by re-prompting the model with the failing test and error message.
We implement our approach in \testpilot, an adaptive LLM-based test generation tool for JavaScript that 
automatically generates unit tests for the methods in a given project's API.

We evaluate \testpilot using OpenAI's \gptturbo LLM on \gptturbonumPackages npm packages with a total of \gptturbonumTotalApiFunctions API functions.
The generated tests achieve a median statement coverage of \gptturbomedianstmtCoverage and branch coverage of \gptturbomedianbranchCoverage.
In contrast, the state-of-the feedback-directed JavaScript test generation technique, Nessie, achieves only \gptturbomedianNessiestmtCoverage statement coverage and \gptturbomedianNessiebranchCoverage branch coverage.
Furthermore, experiments with excluding parts of the information included in the prompts show that all components contribute 
towards the generation of effective test suites. We also find that \gptturbosimLEQfifty 
of \testpilot's generated tests have $\leq$~50\% similarity with existing tests (as measured by normalized edit distance), with none 
of them being exact copies.
Finally, we run \testpilot with two additional LLMs,
OpenAI's older \cushman LLM and \starcoder, an LLM for which the training process is publicly documented. Overall, we
observed similar results with the former 
(\cushmanmedianstmtCoverage median statement coverage), and somewhat worse results with the latter (\starcodermedianstmtCoverage median statement coverage),
suggesting that the effectiveness of the approach is influenced by the size and training set of the LLM, but does not fundamentally depend on the specific model.

\end{abstract}

}

\maketitle

\section{Introduction}

Unit tests check the correctness of individual functions or other units of
source code, and play a key role in modern software
development~\cite{Beck:00,ArtOfAgileBook,TDDBook}. However, creating unit tests
by hand is labor-intensive and tedious, causing some developers to skip writing
tests altogether~\cite{DakaUnitTestPractices2014}.

This fact has inspired extensive research on techniques for automated test
generation including fuzzing~\cite{DBLP:journals/cacm/MillerFS90, AFLSite},
feedback-directed random test generation~\cite{Csallner2004,Pacheco:2007,Pacheco:2008:FEN:1390630.1390643,DBLP:journals/pacmpl/SelakovicPKT18,DBLP:conf/icse/ArtecaHPT22},
dynamic symbolic execution~\cite{DBLP:conf/pldi/GodefroidKS05,DBLP:conf/sigsoft/SenMA05,DBLP:conf/ccs/CadarGPDE06,DBLP:conf/kbse/TillmannHX14},
and search-based and evolutionary techniques~\cite{DBLP:conf/qsic/FraserA11,DBLP:conf/sigsoft/FraserA11}.   
At a high level, most of these techniques use static or dynamic analysis
techniques to explore control and data flow paths in the program, and then
attempt to generate tests that maximize coverage.
While they are often successful in generating tests that expose
faults, these techniques have two major shortcomings. First, the generated tests are
typically less readable and understandable than manually written
tests~\cite{industrialUnitTest,empiricalReadability}, especially due to the use
of unintuitive variable names \cite{DBLP:conf/sigsoft/DakaCFDW15}.
Second, the generated tests often lack
assertions~\cite{DBLP:conf/icsm/PanichellaPFSH20}, or only contain very generic
assertions (e.g.,  that a dereferenced variable must not be \code{null}), or too
many spurious assertions~\cite{DBLP:conf/icse/PalombaNPOL16}. While
such tests can provide inspiration for manually crafting
high-coverage test suites, they do not look natural and generally cannot 
be used verbatim.

Given these disadvantages, there has recently been increasing interest in
utilizing machine learning-based code-generation techniques to produce better
unit tests~\cite{codet,tappy,bareiss22,tufano2021unit,LemieuxICSE2023,atlas,MastrapaoloICSE2021}. Specifically, these
research efforts leverage LLMs that have been
trained on large corpora of natural-language text and source code. We are
specifically interested in \emph{generative transformer models} that, when given
a snippet of text or source code (referred to as the \textit{prompt}), will
predict text that is likely to follow it (henceforth referred to as the
\textit{completion}). It turns out that LLMs are good at producing
natural-looking completions for both natural language and source code, and to
some extent ``understand'' the semantics of natural language and code, based on
the statistical relationships on the likelihood of seeing a particular word in a
given context.  
Some LLMs such as BERT~\cite{devlin2018bert} or GPT-3 \cite{GPT3paper} are
trained purely on text extracted from books and other public sources, while
others like OpenAI Codex~\cite{codex21} and AlphaCode~\cite{AlphaCode} are put
through additional training on publicly available source code to make them
better suited for software development tasks~\cite{DBLP:conf/ijcai/LiWLK18,DBLP:conf/sigsoft/SvyatkovskiyDFS20,DBLP:conf/icse/KarampatsisBRSJ20,DBLP:conf/icse/KimZT021,CopilotSite,DBLP:conf/aaai/GuptaPKS17,DBLP:conf/iclr/HellendoornSSMB20,DBLP:conf/nips/AllamanisJB21,DBLP:journals/pacmpl/PradelS18,DBLP:conf/iclr/AllamanisBK18}.

Given the properties of LLMs, it is reasonable to expect that they may be able to generate natural-looking tests. Not only are they likely to
produce code that resembles what a human developer would write (including, for
example, sensible variable names), but LLMs are also likely to produce tests
containing assertions, simply because most tests in their training set do. Thus,
by leveraging LLMs, one might hope to simultaneously address the two shortcomings of
traditional test-generation techniques. On the other hand, one would perhaps not
expect LLMs to produce tests that cover complex edge cases or exercise unusual
function inputs, as these will be rare in the training data, making LLMs more
suitable for generating regression tests than for bug finding.

There has been some exploratory work on using LLMs for test generation.
For example, Barei{\ss} \emph{et al.}~\cite{bareiss22} evaluate the performance
of Codex for test generation. They
follow a few-shot learning paradigm where their prompt includes the
function to be tested along with an example of another function and an
accompanying test to give the model an idea of what a test should look like.
In a limited evaluation on 18 Java methods, they find that this approach
compares favorably to feedback-directed test generation~\cite{Pacheco:2007}.
Similarly, Tufano et al.'s \textsc{AthenaTest}~\cite{tufano2021unit} generates tests using
a BART transformer model~\cite{lewis2019bart} fine-tuned on a training set of
functions and their corresponding tests. They evaluate on five
Java projects, achieving comparable coverage to EvoSuite~\cite{DBLP:conf/sigsoft/FraserA11}.
While these are promising early results, these approaches, as well as others~\cite{MastrapaoloICSE2021,MastrapaoloTSE2022,TufanoAST22Assert}, rely on a training corpus of functions and their corresponding tests, which is expensive to curate
and maintain.

In this paper, we explore the feasibility of automatically generating unit
tests using off-the-shelf LLMs, with no additional training and as little
pre-processing as possible. Following Reynolds and
McDonell~\cite{reynoldsPromptProgramming}, we posit that providing the model
with input-output examples or performing additional training is not necessary and that careful prompt crafting is sufficient.
Specifically, apart from test scaffolding code, our prompts contain (1) the
signature of the function under test; (2) its documentation comment, if any; (3) usage examples for the function mined from documentation, if available; (4) its
source code. Finally, we consider an adaptive component to our technique: each
generated test is executed, and if it fails, the LLM is prompted again with a
special prompt including (5) the failing test and the error message it produced,
which often allows the model to fix the test and make it pass.

To conduct experiments, we have implemented these techniques in a system called \testpilot, an LLM-based test 
generator for JavaScript. %
We chose JavaScript as an example of a popular language
for which test generation using traditional methods is challenging due to the
absence of static type information and its permissive runtime
semantics~\cite{DBLP:conf/icse/ArtecaHPT22}. We evaluate our approach on
\gptturbonumPackages npm packages from various domains hosted on both GitHub and GitLab, with varying levels of popularity and amounts of available documentation. 
These packages have a total of \gptturbonumTotalApiFunctions API functions that we attempt to generate tests for.
We investigate the coverage achieved by the
generated tests and their quality in terms of success rate, reasons for failure,
and whether or not they contain assertions that actually exercise
functionality from the target package (\textit{non-trivial assertions}). We also empirically evaluate the
effect of the various components of our prompt-crafting strategy as well as whether \testpilot is generating previously memorized 
tests from the LLM's training data.

Using OpenAI's current most capable and cost-effective model \gptturbo,\footnote{\url{https://platform.openai.com/docs/models/gpt-3-5}} \testpilot's generated tests achieve a median statement coverage of \gptturbomedianstmtCoverage, and branch coverage of \gptturbomedianbranchCoverage.
We find that a median \gptturbomedianPercentNonTrivialTests of the generated tests contain non-trivial assertions, and that these non-trivial tests alone achieve a median \gptturbomediannonTrivialCoverage coverage, indicating that the generated tests contain meaningful oracles that exercise functionality from the target package.
Upon deeper examination, we find that the most common reason for the generated tests to fail is exceeding the two-second timeout we enforce,
usually because of a failure to communicate test completion to the testing framework.
We find that, on average, the adaptive approach is able to fix \gptturboallErrorsFixedWithRetry of failing tests. 
Our empirical evaluation also shows that all five components included in the prompts are essential for generating meaningful test suites with high coverage.
Excluding any of these components results in either a higher proportion of failing tests or in reduced coverage.
On the other hand, while excluding usage examples from prompts reduces effectiveness of the approach, it does not render
it obsolete, suggesting that the LLM is able to learn from the presence of similar test code in its training set.   

Finally, from experiments conducted with the \gptturbo LLM, we note that high coverage is still achieved on packages whose source code 
is hosted on GitLab (and thus has not been part of the LLM's training data).
Moreover, we find that \gptturbosimLEQforty of the tests generated using the \gptturbo LLM have $\leq$ 40\% similarity to existing tests and 
\gptturbosimLEQfifty have $\leq$ 50\% similarity, with none of the tests being exact copies. This suggests that the generated tests are not
copied verbatim from the LLM's training set.

In principle, the test generation approach under consideration can be used with any LLM. However, the effectiveness of the approach
is likely to depend on the LLM's size and training set. To explore this factor, we further conducted experiments with two additional LLMs: the previous proprietary \cushman~\cite{CushmanSite} model developed by OpenAI and \starcoder~\cite{StarCoderSite}, 
an LLM for which the training process is publicly documented.
We observed qualitatively similar results using \cushman 
(median coverage of \cushmanmedianstmtCoverage for statements, \cushmanmedianbranchCoverage for branches), and somewhat worse results using \starcoder
(\starcodermedianstmtCoverage and \starcodermedianbranchCoverage).

In summary, this paper makes the following contributions:
\begin{itemize}[leftmargin=*]
  \item
    A simple test generation technique where unit tests are generated by iteratively querying an LLM with a prompt
    containing signatures of API functions under test and, optionally, the bodies, documentation, and usage examples associated 
    with such functions. The technique also features an adaptive component that includes in a prompt error messages observed when
    executing previously generated tests. 
  \item
    An implementation of this technique for JavaScript in a tool called \testpilot, which is available as open-source software at \url{https://github.com/githubnext/testpilot}.
  \item
    An extensive empirical evaluation of \testpilot on \gptturbonumPackages npm packages, demonstrating its effectiveness in generating test suites 
    with high coverage. Our evaluation explores the following aspects:
	\begin{itemize}
		\item Quality of the generated tests in terms of the assertions they contain, and coverage of tests that include non-trivial assertions.
		\item Effect of excluding various prompt components.
		\item Similarity of generated tests to existing tests.
		\item Comparison against Nessie \cite{DBLP:conf/icse/ArtecaHPT22}, a state-of-the-art feedback-directed random test generation technique for JavaScript.
		\item Comparison of the effect of the underlying LLM on \testpilot's generated tests.
	\end{itemize}
\end{itemize}

The raw data and analysis for all our experiments can be found at \url{https://doi.org/10.6084/m9.figshare.23653371}.
\section{Approach}
\label{sec:approach}

\testpilot{} generates tests using the popular JavaScript testing framework
Mocha~\cite{mocha} with its BDD-style syntax in which tests are implemented
as callback functions that are passed  to the \code{it} function. Test suites
consist of one or more calls to  \code{it} that occur in a callback function that
is passed to the \code{describe} function. Assertions are checked using the built-in
Node.js \code{assert} module. 

\begin{figure}
\vspace*{2mm}
\begin{minted}[fontsize=\small,linenos,highlightlines={1-8},mathescape]{javascript}
let mocha = require('mocha');
let assert = require('assert');
let pkg = require('package-under-test');

// function metadata, including signature of f

describe('test pkg', function() {
  it('test f', function(done) {
    // test code, terminated by done() $\label{line:testcode}$
  })
})
\end{minted}
\caption{Illustration of the structure of prompts and tests.}
\label{fig:ExamplePrompt}
\end{figure}

Figure~\ref{fig:ExamplePrompt} illustrates the structure of generated tests for a 
function $f$. Here, lines 1--3 are boilerplate code for importing the testing libraries and the
Package under Test (PUT).  These are followed by one or more commented-out lines
containing function metadata included in the prompt, as we explain shortly.
Lines 7--8 begin the definition of a test suite using \code{describe} with a
single test defined as a callback function accepting a parameter \code{done}
passed to the \code{it} function.  The test code uses \code{assert} to check its
assertions, and finally invokes \code{done()} to signal completion. This is
necessary for asynchronous tests that may take multiple iterations of the
JavaScript event loop to finish. Calling \code{done()} more than once results in
a runtime error, while not calling it at all causes the test to fail with a
timeout error.

The basic idea of our approach is to send the initial part of the above
test skeleton up to (but not including) the start of the actual test code on
Line~\ref{line:testcode} (highlighted above in blue) as a prompt to the LLM. Since LLMs
are trained to complete a given code fragment, one might therefore expect it to
generate the rest of the test for us. Comments can be included in the test
skeleton to provide additional information about the function that may be useful
to guide the LLM towards generating better tests.

\subsection{\testpilot Architecture}

\begin{figure*}
\centering
  \includegraphics[width=0.8\textwidth]{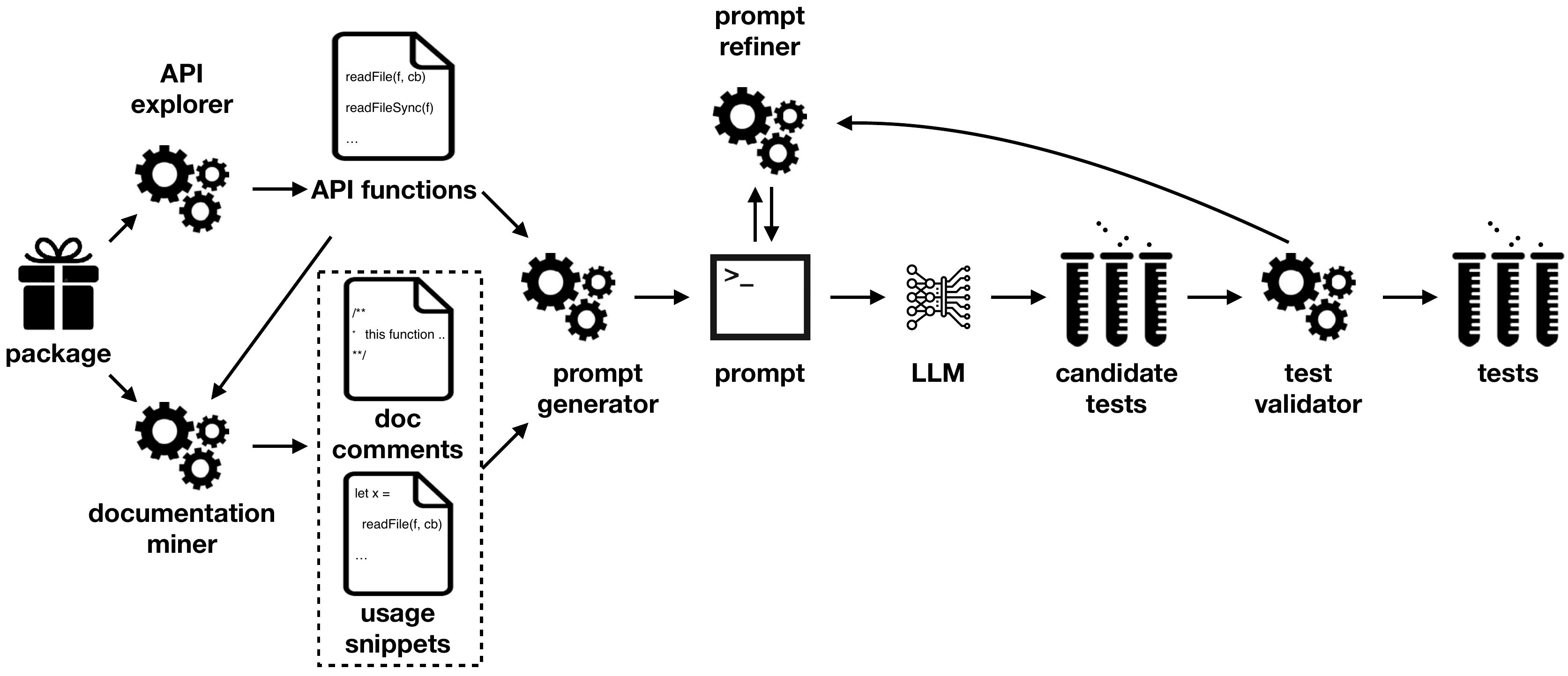}
  \caption{Overview of the adaptive test generation technique we use in \testpilot.\vspace{-0.4cm}}
  \label{fig:overview}
\end{figure*} 

Figure~\ref{fig:overview} presents the high-level architecture of \testpilot, which consists of five main components:
Given a PUT as input, the \textit{API explorer} identifies
functions to test; the \textit{documentation miner} extracts metadata about them; and the
\textit{prompt generator}, \textit{test validator}, and \textit{prompt refiner} collaborate to construct
prompts for test generation, assemble complete tests from the LLM's response,
run them to determine whether they pass, and construct further prompts to
generate more tests.
We now discuss each of these components in more detail.

\paragraph*{\bf API Explorer} 
This component analyzes the PUT to determine its API, i.e., the set of functions, methods, constants, etc. that the package exposes
to clients.  In JavaScript, it is very difficult to determine the API statically
due to the highly dynamic nature of the language. Therefore, 
similar to other JavaScript test-generation work~\cite{DBLP:journals/pacmpl/SelakovicPKT18,DBLP:conf/icse/ArtecaHPT22},
we pursue an approach based on dynamic analysis. In particular, we
load the application's main package and apply introspection to traverse the resulting
object graph and identify properties that are bound to functions. For each function, we record its access
path (that is, the sequence of properties that must be traversed to reach it
from the main module), its signature (which in the absence of static type
information is simply a list of parameter names), and its definition (that is,
its source code). The output of the API Explorer is a list of functions
described by their access paths, signatures, and definitions; other API elements
are ignored.

\paragraph*{\bf Documentation Miner}
This component extracts code snippets and comments from documentation included with the PUT,
and associates them with the API functions they pertain to. The aim is to collect, for each
API function, comments and examples describing its purpose and intended usage.
In JavaScript code bases, documentation is typically provided in the form of
Markdown (\code{.md}) files, in which code snippets are embedded as fenced code
blocks (i.e., blocks surrounded by triple backticks).  We find all such blocks in all Markdown files in the code base, and
associate with each function the set of all code snippets that textually contain
the function's name.  While this is a simple heuristic, code examples may not be complete or syntactically correct, so a more
sophisticated approach relying on parsing or static analysis is not likely to
work well.
We also associate each API function with the doc comment (\texttt{/**\ldots*/}) that immediately precedes it,
if any.

\begin{figure*}
\centering
\begin{tabular}{c@{\hspace*{0.05\textwidth}}c}
  \begin{minipage}[t]{0.45\textwidth}
    \begin{minted}[linenos,mathescape,fontsize=\footnotesize,highlightlines={1-6}]{javascript}
let mocha = require('mocha');
let assert = require('assert');
let countries_and_timezones = require('countries-and-timezones');
// countries-and-timezones.getCountry(id)
describe('test countries_and_timezones', function() {
  it('test countries-and-timezones.getCountry', function(done) {
    let country = countries_and_timezones.getCountry('US');
    assert.equal(country.name, 'United States'); // fails $\label{line:FailingAssertion}$
    assert.equal(country.timezones.length, 2);
    assert.equal(country.timezones[0], 'America/New_York');
    assert.equal(country.timezones[1], 'America/Chicago');
    done();
  })
})
    \end{minted}
  \end{minipage}
&
  \begin{minipage}[t]{0.45\textwidth}
    \begin{minted}[linenos,fontsize=\footnotesize,highlightlines={1-22}]{javascript}
let mocha = require('mocha');
let assert = require('assert');
let countries_and_timezones = require('countries-and-timezones');
// usage #1
// const ct = require('countries-and-timezones');
// 
// const country = ct.getCountry('DE');
// console.log(country);
// 
// /*
// Prints:
// 
// {
//   id: 'DE',
//   name: 'Germany',
//   timezones: [ 'Europe/Berlin', 'Europe/Zurich' ]
// }
// 
// */
// countries-and-timezones.getCountry(id)
describe('test countries_and_timezones', function() {
  it('test countries-and-timezones.getCountry', function(done) {
    let country = countries_and_timezones.getCountry('DE');
    assert.equal(country.id, 'DE');
    assert.equal(country.name, 'Germany');
    assert.equal(country.timezones[0], 'Europe/Berlin');
    assert.equal(country.timezones[1], 'Europe/Zurich');
    done();
  })
})
      \end{minted}
  \end{minipage}
  \\
  {\bf (a)} & {\bf (b)}
\end{tabular}
\caption{Examples of prompts (highlighted) and the completions provided
  by the LLM, comprising complete tests. 
  Prompt~(a) contains no snippets and the test generated from it fails.
  Prompt~(b) contains one snippet and the generated test passes. \vspace{-0.5cm}
}
  \label{fig:PromptExamples}
\end{figure*}

\vspace*{2mm}
The remaining three components are the \textit{prompt generator}, the \textit{test validator}, and the
\textit{prompt refiner}, which work together to generate and validate tests for all 
API functions identified by the API Explorer, using the information provided by
the Documentation Miner.  Functions are processed one at a time, and for each function
only one test is generated at a time (as opposed to generating an entire test
suite at once).  This is to enable us to validate each test individually
without interference from other tests.

\paragraph*{\bf Prompt Generator}
This component constructs the initial prompt to send to the LLM for generating
a test for a given function $f$.  As mentioned above, we initially have (at
most) four pieces of information about $f$ at our disposal: its signature, its
definition, its doc comment, and its usage snippets extracted from documentation.  While it might seem natural
to construct a prompt containing all of this information, in practice it can
sometimes happen that more complex prompts lead to worse completions as the
LLM gets confused by the additional information.  Therefore, we follow a
different strategy: we start with a very simple initial prompt that includes no
metadata except the function signature, and then let the prompt refiner extend
it step by step with additional information.
 
\paragraph*{\bf Test Validator}
Next, we send the generated prompts to LLM and wait for completions.
We only consume as many tokens as are needed to form a syntactically valid test. Since there is no guarantee that the completions suggested by the model are
syntactically valid, the test validator tries to fix simple syntactic
errors such as missing brackets, and then parses the resulting code to check
whether it is syntactically valid.  If not, the test is immediately marked as
failed. Otherwise it is run using the Mocha test runner to determine whether it
passes or fails (either due to an assertion error or some other runtime error).
 
Each returned completion can be
concatenated with the prompt to yield a candidate test. 
However, to allow us to eliminate duplicate tests generated from different prompts, we post-process the candidate tests as follows:
we strip out the comment containing the function metadata in the prompt and replace
the descriptions in the \code{describe} and \code{it} calls with the generic
strings \code{'test suite'} and \code{'test case'}, respectively.

\paragraph*{\bf Prompt Refiner}
The Prompt Refiner applies a number of strategies to generate additional prompts to use for querying the model.
Overall, we employ four prompt refiners as follows:

\begin{enumerate}[leftmargin=*]
  \item \FunctionBodyIncluder: If $p$ did not contain the definition of $f$, a prompt is created that includes it.
  \item \DocCommentIncluder: If $f$ has a doc comment but $p$ did not include it, a prompt with the doc comment is created.
  \item \SnippetIncluder: If usage snippets for $f$ are available but $p$ did not include them, a prompt with snippets is created.
  \item \RetryWithError: If $t$ failed with error message $e$, a prompt is constructed that consists of:
      the text of the failing test $t$ followed by 
      a comment \texttt{// the test above fails with the following error: $e$}, 
      followed by a comment \texttt{// fixed test}. This strategy is only applied once per
      prompt, so it is not attempted if $p$ itself was already generated by this strategy.
\end{enumerate}  

The refined prompt is then used to construct a test in the same way as the
original prompt.  All strategies are applied independently and in all possible
combinations, but note that the first three will only apply at most once and the
fourth will never apply twice in a row, thus ensuring termination.

\begin{algorithm}[tb]
\caption{Pseudo-code for API exploration.}
\label{alg:exploreAPI}
{\footnotesize
\begin{algorithmic}[1]

\Function{\textit{exploreAPI}}{$\pkgName$}
\State \modObj $\leftarrow$ object created by importing $\pkgName$ \label{line:createModObj}  
\State \seen $\leftarrow$ $\emptyset$
\State {\bf return} \explore{\pkgName}{\modObj}{\seen}
\EndFunction

\Function{\textit{explore}}{$\accessPath,\obj,\seen$}
\State \apis $\leftarrow$ $\emptyset$
\If{\obj $\not\in$ \seen} %
  \State \seen $\leftarrow$ \seen $\cup$ $\{$ \obj $\}$
  \If{\obj is a function with signature \sig} 
    \State \apis $\leftarrow$ \apis $\cup$ $\Set{ \Pair{\accessPath}{\sig} }$ \label{line:RecordFunctionProperty}
  \ElsIf{\obj is an object} 
    \State \props $\leftarrow$  $\{$ \prop $|$ \obj has a property \prop $\}$ %
    \For{\prop in \props}
       \State \apis $\leftarrow$ \apis $\cup$ \\ \label{line:ExploreObjectPropsStart}
             \hspace*{2cm} \explore{\extend(\accessPath, \prop)}{$\obj{}$[\prop]}{\seen} 
    \EndFor \label{line:ExploreObjectPropsEnd}
   \ElsIf{obj is an array}
     \For{each index $i$ in the array} \label{line:ExploreArrayElementsStart}
       \State \apis $\leftarrow$ \apis $\cup$ \\ 
              \hspace*{2cm} \explore{\extend(\accessPath, \prop)}{$\obj{}$[i]}{\seen} \label{line:ExploreArrayElementsEnd}
     \EndFor
   \EndIf
\EndIf
\State \textbf{return} \apis
\EndFunction

\Function{\textit{extend}}{$\accessPath,\component$}
\If{\component is numeric}
\State \textbf{return} $\accessPath{}$[\component]
\Else
\State \textbf{return} $\accessPath{}$.\component
\EndIf
\EndFunction

\end{algorithmic}
}
\end{algorithm} 

\subsection{Algorithm Details}

We now provide additional detail on the two key steps of our approach: API exploration and test generation.  

\paragraph*{\bf API Exploration}

Algorithm~\ref{alg:exploreAPI} shows pseudocode that illustrates how the set of functions that constitute the API for a package
is identified. The algorithm takes a package under test, \pkgName, and produces a list of pairs $\Pair{a}{\sig}$ representing its API.
Here, $a$ is an \textit{access path} that uniquely represents an API method, and \sig is the signature of a function.
Our notion of an access path takes a somewhat simplified form compared to the original concept proposed by
Mezzetti et al. \cite{mezzetti18}, and consists of a package name followed by a sequence of property names.        

We rely on a dynamic approach to explore the API of a package \pkgName, by creating a small program that imports the package 
(line~\ref{line:createModObj}), and relying on JavaScript's introspective capabilities to determine which properties are present 
in the \textit{package root} object \modObj that is created by importing \pkgName and what types these properties have. 
Exploration of  \modObj's properties is handled by a recursive function \textit{explore} that begins at the access path
representing the package root and that traverses this object recursively,  calling another auxiliary function \textit{extend} 
to extend the access path as the traversal descends into the object's structure.   

During exploration, if an object is encountered at access path $a$  whose type is a function with signature \sig, then a 
pair $\Pair{a}{\sig$} is recorded (line~\ref{line:RecordFunctionProperty}). If the type of $p$ is an object,
then the objects referenced by its properties are recursively explored (lines~\ref{line:ExploreObjectPropsStart}--\ref{line:ExploreObjectPropsEnd}), 
and if the type of $p$ is an array, then $p$'s properties are explored recursively as well (lines~\ref{line:ExploreArrayElementsStart}--\ref{line:ExploreArrayElementsEnd}).

\begin{algorithm*}[tb]
\caption{Pseudo-code for test generation.}
\label{alg:TestGeneration}
{\footnotesize
\begin{algorithmic}[1]
	\Function{\textit{generateTests}}{\pkgName, \LLM}	
	\State \prompts $\leftarrow$ $\emptyset$ \label{line:InitializePrompts}
	\State \tests $\leftarrow$ $\emptyset$  \label{line:InitializeTests}
	\State \seen $\leftarrow$ $\emptyset$  \label{line:InitializeSeen}
	\State \apis $\leftarrow$ \exploreAPI{\pkgName} \label{line:callExploreAPI}  \Comment{See Algorithm~\ref{alg:exploreAPI}}
	\For{$\api \in \apis$} \label{line:createBasePromptsStart}
	  \State \prompts $\leftarrow$ \prompts $\cup$ \Set{ \createBasePrompt{\api} } \Comment{create base prompts containing only the signature, see Figure~\ref{fig:ExamplePrompt}}
	\EndFor \label{line:createBasePromptsEnd}
	\State
	
	\State \promptsWithFnBody $\leftarrow$ $\emptyset$       \Comment{refine prompts by adding function body} \label{line:StartRefining}
	\For{$\prompt \in \prompts$} 
	  \State \body $\in$ \findFunctionBodyForSig{\prompt.\api.\accessPath}{\prompt.\api.\sig} 
	  \State \promptsWithFnBody $\leftarrow$ \promptsWithFnBody $\cup$ \refine{\prompt}{\body} 
	\EndFor  
	\State \prompts $\leftarrow$ \prompts $\cup$ \promptsWithFnBody
	\State  
	
	\State \promptsWithExampleSnippets $\leftarrow$ $\emptyset$  \Comment{refine prompts in cases where example snippets are available}
	\For{$\prompt \in \prompts$}  
  	  \State \snippets $\leftarrow$ \findExamplesForSig{\prompt.\api.\accessPath}{\prompt.\api.\sig}
	  \If{\snippets $\neq$ $\emptyset$} 
	    \State \promptsWithExampleSnippets $\leftarrow$ \promptsWithExampleSnippets $\cup$ \refine{\prompt}{\snippets}
	  \EndIf
	\EndFor
	\State \prompts $\leftarrow$ \prompts $\cup$ \promptsWithExampleSnippets
	\State
	
	\State \promptsWithDocComments $\leftarrow$ $\emptyset$ \Comment{refine prompts in cases where doc comments are available}   
	\For{$\prompt \in \prompts$} 
	  \State \docComment $\leftarrow$ \findDocCommentsForSig{\prompt.\api.\accessPath}{\prompt.\api.\sig}
	  \If{\docComment $\neq$ $\emptyset$}
  	    \State \promptsWithDocComments $\leftarrow$ \promptsWithDocComments $\cup$ \refine{\prompt}{\docComment} 
	  \EndIf
	\EndFor 
	\State \prompts $\leftarrow$ \prompts $\cup$ \promptsWithDocComments \label{line:EndRefining}

	\State
	\While{\prompts $\neq$ $\emptyset$} \label{line:WhileLoopStart}
	  \State select and remove \prompt from \prompts
	  \State \completions $\leftarrow$ \getCompletions{\LLM}{\prompt.\textit{text}} \label{line:queryLLM} \Comment{request completions from the LLM}
	  \For{\completion $\in$ \completions}
	    \State \test $\leftarrow$ \concatenate{prompt}{completion} \label{line:concatenate}
	    \State \test $\leftarrow$ \fixMinorSyntaxIssues{test} \Comment{e.g., add missing close parentheses} \label{line:fixSyntax}
	    \State \test $\leftarrow$ \removeComments{test} \label{line:removeComments}
 	    \If{\test is syntactically valid \textbf{and} \test $\not\in$ \seen}
 	      \State \seen $\leftarrow$ \seen $\cup$ \Set{ \test }
 	      \State \result $\leftarrow$ \executeTest{\test} \Comment{execute the test} \label{line:executeTest}
	      \If{\resultStatus = \ok} \Comment{add successful test to \tests}
	          \State \tests $\leftarrow$ \tests $\cup$ \Set{ \test } \label{line:addPassingTest}
	      \Else \Comment{\resultStatus = \assertionFailure \textbf{or} \resultStatus = \crash \textbf{or} \resultStatus = \nonTermination} 
	        \If{\prompt was not constructed from a previous failed test} \Comment{apply error retry refiner} \label{line:fixFailingTest}
	          \State \newPrompt $\leftarrow$ \refineFromError{\test}{\resultErrorMessage} 
	          \State \prompts $\leftarrow$ \prompts $\cup$ \Set{ \newPrompt } 
	       \EndIf 
	      \EndIf
       \EndIf
	  \EndFor
	\EndWhile \label{line:WhileLoopEnd}
	\State \Return{\tests} \label{line:returnTests}
    \EndFunction
\end{algorithmic}
}
\end{algorithm*}

\paragraph*{\bf Test Generation}

Algorithm~\ref{alg:TestGeneration} shows pseudo-code for the test generation step.
The algorithm begins by initializing the set \prompts of generated prompts, the set \tests of generated passing tests, and the set \seen
containing all generated tests to the empty set and by using Algorithm~\ref{alg:exploreAPI} to obtain the set \apis of (access path, signature) pairs that constitute the 
package's API (lines~\ref{line:InitializePrompts}--\ref{line:callExploreAPI}).
Then, on lines~\ref{line:createBasePromptsStart}--\ref{line:createBasePromptsEnd}, for each such pair, a base prompt is 
constructed and added to \prompts, containing only the access path and signature, using the template illustrated in Figure~\ref{fig:ExamplePrompt}.
Next, lines~\ref{line:StartRefining}--\ref{line:EndRefining} create additional prompts by adding the function body, example usage snippets, and
documentation comments extracted from the code to previously generated prompts.  Here, the \textit{refine} function extends a previously
generated prompt by adding the function body, example snippets, or doc comment.
The order in which each type of information, if included, appears in prompts is fixed as follows:
    example snippets,
    error message from previously generated test,
    doc comments,
    function body,
    signature.

The while loop on lines~\ref{line:WhileLoopStart}--\ref{line:WhileLoopEnd} describes an iterative process for generating tests that continues
as long as prompts remain that have not been processed. In each iteration, a prompt is selected and removed from \prompts, and the   
LLM is queried for completions (line~\ref{line:queryLLM}).
For each \completion that was received, a test is constructed by concatenating the prompt and the completion (line~\ref{line:concatenate})
and minor syntactic problems are fixed such as adding missing `\texttt{\}}' characters at the end of the test (line~\ref{line:fixSyntax}).
Moreover, we remove comments from the test to enable deduplication of tests that only differ in their comments (line~\ref{line:removeComments}).

If the resulting test is syntactically valid and the same test was not encountered previously, it is executed (line~\ref{line:executeTest}).
Otherwise, we do not re-execute it but still link the prompt to the previously seen test. 
If the test executed successfully, we add it to \tests (line~\ref{line:addPassingTest}).
If it failed (due to an assertion failure, nontermination, or because of an uncaught exception), and if the test was not derived from a prompt that 
was constructed from a previous failing test
(line~\ref{line:fixFailingTest}), then we create a new prompt containing the failed test and the error message and add it to \prompts.

When the iterative process concludes, the set \tests is returned (line~\ref{line:returnTests}).

\begin{figure*}
\centering
\begin{tabular}{c@{\hspace*{0.05\textwidth}}c}
  \begin{minipage}[t]{0.45\textwidth}
    \begin{minted}[linenos,mathescape,fontsize=\footnotesize,highlightlines={1-6}]{javascript}
let mocha = require('mocha');
let assert = require('assert');
let quill_delta = require('quill-delta');
// quill-delta.prototype.concat(other)
describe('test quill_delta', function() {
  it('test quill-delta.prototype.concat', function(done) {
    let delta1 = new quill_delta([{ insert: 'Hello' }, 
                                  { insert: ' ', 
                                    attributes: { bold: true } }, 
                                  { insert: 'World!' }]);
    let delta2 = new quill_delta([{ insert: 'Hello' }, 
                                  { insert: ' ', 
                                    attributes: { bold: true } }, 
                                  { insert: 'World!' }]);
    let delta3 = delta1.concat(delta2);
    assert.equal(delta3.ops.length, 6); // fails $\label{line:FailingAssertionRefiner}$
    done();
  })
})
    \end{minted}
  \end{minipage}
&
  \begin{minipage}[t]{0.45\textwidth}
    \begin{minted}[linenos,fontsize=\footnotesize,highlightlines={1-23}]{javascript}
let mocha = require('mocha');
let assert = require('assert');
let quill_delta = require('quill-delta');
// quill-delta.prototype.concat(other)
describe('test quill_delta', function() {
  it('test quill-delta.prototype.concat', function(done) {
    let delta1 = new quill_delta([{ insert: 'Hello' }, 
                                  { insert: ' ', 
                                    attributes: { bold: true } }, 
                                  { insert: 'World!' }]);
    let delta2 = new quill_delta([{ insert: 'Hello' }, 
                                  { insert: ' ', 
                                    attributes: { bold: true } }, 
                                  { insert: 'World!' }]);
    let delta3 = delta1.concat(delta2);
    assert.equal(delta3.ops.length, 6);
    done();
  })

  // the test above fails with the following error:
  //   expected 5 to equal 6
  // fixed test:
  it('test quill_delta', function(done) {
    let delta1 = new quill_delta([{ insert: 'Hello' }, 
                                  { insert: ' ', 
                                    attributes: { bold: true } }, 
                                  { insert: 'World!' }]);
    let delta2 = new quill_delta([{ insert: 'Hello' }, 
                                  { insert: ' ', 
                                    attributes: { bold: true } }, 
                                  { insert: 'World!' }]);
    let delta3 = delta1.concat(delta2);
    assert.equal(delta3.ops.length, 5);
    done();
  })
}
  \end{minted}
  \end{minipage}
  \\
  {\bf (a)} & {\bf (b)}
\end{tabular}
\caption{ 
  Example illustrating how a prompt is refined in response to the failure of a previously generated test.
  Prompt (a) contains no information except the method signature, and the test generated from it fails.
  Prompt (b) adds information about the test failure, and the generated test passes.\vspace{-0.4cm}
}
  \label{fig:RefinerExample}
\end{figure*}

\subsection{Examples}

To make the discussion more concrete, we will now show two examples of how
\testpilot{} generates tests.

\begin{sloppypar}
As the first example, we consider the npm package
\texttt{countries-and-timezones}.%
\footnote{See \url{https://www.npmjs.com/package/countries-and-timezones}.}  
API exploration reveals that this package exports a function \code{getCountry}
with a single parameter \code{id} and the project's \code{README.md} file
provides a usage example.
\end{sloppypar}

Figure~\ref{fig:PromptExamples}(a) shows a test for this function generated from
the initial highlighted prompt that only includes the function signature, but
no other metadata. This test fails when execution reaches the assertion on
line~\ref{line:FailingAssertion} because the expression \code{country.name}
evaluates to \mintinline{javascript}{"United States of America"}, which differs
from the value \mintinline{javascript}{"United States"} expected by the
assertion.

Next, we refine this prompt to include the usage snippet as shown in the highlighted part of
Figure~\ref{fig:PromptExamples}(b).  This enables the LLM to generate a test
incorporating the information provided in this snippet, which passes when
executed.

We show another example in Figure~\ref{fig:RefinerExample} from \texttt{quill-delta},%
\footnote{ See \url{https://github.com/quilljs/delta}. } a package for
representing and manipulating changes to documents.
As before, Figure~\ref{fig:RefinerExample}(a) shows the initial prompt for
\texttt{quill-delta}'s \code{concat} method, which concatenates two change sets,
and a test that was generated from this prompt.   It is noteworthy that the LLM
was able to generate a syntactically correct test for \texttt{quill-delta},
where arguments such as 
\begin{minted}[linenos,fontsize=\footnotesize]{javascript}
 [{ insert: 'Hello' }, 
  { insert: ' ', attributes: { bold: true } }, 
  { insert: 'World!' }]
\end{minted}
 are passed to the constructor \textit{even in the absence of any usage examples}.  
Most likely, this is because \texttt{quill-delta} is a popular package with more
than 1.2 million weekly downloads, which means that the LLM is likely to
have seen examples of its use in its training set.

Nevertheless, the test in Figure~\ref{fig:RefinerExample}(a) fails because when
reaching the assertion on line~\ref{line:FailingAssertionRefiner}
\code{delta3.ops.length} has the value \code{5}, whereas the assertion expects
the value \code{6}. The reason for the assertion's failure is the fact that the
\code{concat} method merges adjacent elements if they have the same attributes.
Therefore, when execution reaches line~\ref{line:FailingAssertionRefiner}, the
array \code{delta3.ops} will hold the following value:
 \begin{minted}[linenos,fontsize=\footnotesize]{javascript}
[
  { insert: 'Hello' },
  { insert: ' ', attributes: { bold: true } },
  { insert: 'World!Hello' },
  { insert: ' ', attributes: { bold: true } },
  { insert: 'World!' }
]
\end{minted}
and therefore \code{delta3.ops.length} will have the value \code{5}.

\begin{table*}[t!]
\centering
\caption{Overview of npm packages used for evaluation, ordered by descending popularity in terms of downloads/wk. The top 10 packages correspond to the Nessie benchmark, the next 10 are additional GitHub-hosted packages we include, while the last 5 are GitLab-hosted packages.}
\label{tab:packages}
\resizebox{0.8\textwidth}{!}{
  \begin{tabular}{@{}llrrrrrrr@{}}
  \toprule                                                                                                                            
  \multirow{2}{*}{\textbf{Package}}  & 
  \multirow{2}{*}{\textbf{Domain}} & 
  \multirow{2}{*}{\textbf{LOC}} & 
  \multirow{2}{*}{\thead{\textbf{Existing}\\ \textbf{Tests}}} &
  \multirow{2}{*}{\thead{\textbf{Weekly}\\ \textbf{Downloads}}} & 
  \multicolumn{3}{c}{\textbf{API functions}} & 
  \multirow{2}{*}{\thead{\textbf{Total}\\ \textbf{Examples}}}\\ 
  \cmidrule{6-8}
  &&&&& \textbf{\#} & \textbf{\# (\%) w/ examples} & \textbf{\# (\%) w/ comment}\\                                                                                                                                                                                                                                                 
  \midrule
 \href{https://github.com/isaacs/node-glob/commit/8315c2d576f9f3092cdc2f2cc41a398bc656035a}{\color{blue} glob} & 
    file system & 
    314 & 
    22 &
    103M & 
    21 & 
    2 (9.5\%) & 
    0 (0.0\%) & 
    4\\ 
\href{https://github.com/jprichardson/node-fs-extra/commit/6bffcd81881ae474d3d1765be7dd389b5edfd0e0}{\color{blue} fs-extra} & 
    file system & 
    822 & 
    417 &
    79M & 
    172 & 
    23 (13.4\%) & 
    0 (0.0\%) & 
    27\\ 
\href{https://github.com/isaacs/node-graceful-fs/commit/c1b377782112ae0f25b2abe561fbbea6cfb6f876}{\color{blue} graceful-fs} & 
    file system & 
    208 & 
    11 &
    48M & 
    137 & 
    1 (0.7\%) & 
    0 (0.0\%) & 
    1\\ 
\href{https://github.com/jprichardson/node-jsonfile/commit/9c6478a85899a9318547a6e9514b0403166d8c5c}{\color{blue} jsonfile} & 
    file system & 
    46 & 
    43 &
    48M & 
    4 & 
    4 (100.0\%) & 
    0 (0.0\%) & 
    14\\ 
\href{https://github.com/petkaantonov/bluebird/commit/6c8c069c34829557abfaca66d7f22383b389a4b5}{\color{blue} bluebird} & 
    promises & 
    3.1K & 
    238 &
    26M & 
    115 & 
    59 (51.3\%) & 
    0 (0.0\%) & 
    248\\ 
\href{https://github.com/kriskowal/q/commit/6bc7f524eb104aca8bffde95f180b5210eb8dd4b}{\color{blue} q} & 
    promises & 
    736 & 
    214 &
    14M & 
    98 & 
    29 (29.6\%) & 
    15 (15.3\%) & 
    64\\ 
\href{https://github.com/tildeio/rsvp.js/commit/21e0c9720e08ffa53d597c54fed17119899a9a83}{\color{blue} rsvp} & 
    promises & 
    565 & 
    171 &
    8.6M & 
    29 & 
    11 (37.9\%) & 
    16 (55.2\%) & 
    15\\ 
\href{https://github.com/streamich/memfs/commit/ec83e6fe1f57432eac2ab61c5367ba9ec3a775a1}{\color{blue} memfs} & 
    file system & 
    2.2K & 
    265 &
    13M & 
    376 & 
    21 (5.6\%) & 
    7 (1.9\%) & 
    26\\ 
\href{https://github.com/fshost/node-dir/commit/a57c3b1b571dd91f464ae398090ba40f64ba38a2}{\color{blue} node-dir} & 
    file system & 
    244 & 
    55 &
    6M & 
    6 & 
    6 (100.0\%) & 
    5 (83.3\%) & 
    8\\ 
\href{https://github.com/maugenst/zip-a-folder/commit/5089113647753d5086ea20f052f9d29840866ee1}{\color{blue} zip-a-folder} & 
    file system & 
    25 & 
    5 &
    95K & 
    3 & 
    2 (66.7\%) & 
    0 (0.0\%) & 
    2\\ 
\midrule 
\href{https://github.com/js-sdsl/js-sdsl/commit/055866ad5515037c724a529fecb2d3c2b35b2075}{\color{blue} js-sdsl} & 
    data structures & 
    1.5K & 
    88 &
    9.7M & 
    133 & 
    3 (2.3\%) & 
    0 (0.0\%) & 
    1\\ 
\href{https://github.com/quilljs/delta/commit/5ffb853d645aa5b4c93e42aa52697e2824afc869}{\color{blue} quill-delta} & 
    document changes & 
    395 & 
    180 &
    1.6M & 
    36 & 
    17 (47.2\%) & 
    0 (0.0\%) & 
    17\\ 
\href{https://github.com/infusion/Complex.js/commit/d995ca105e8adef4c38d0ace50643daf84e0dd1c}{\color{blue} complex.js} & 
    numbers/arithmetic & 
    393 & 
    21 &
    497K & 
    52 & 
    7 (13.5\%) & 
    52 (100.0\%) & 
    5\\ 
\href{https://github.com/pull-stream/pull-stream/commit/29b4868bb3864c427c3988855c5d65ad5cb2cb1c}{\color{blue} pull-stream} & 
    streams & 
    308 & 
    31 &
    78K & 
    24 & 
    7 (29.2\%) & 
    0 (0.0\%) & 
    7\\ 
\href{https://github.com/manuelmhtr/countries-and-timezones/commit/e34cb4b6832795cbac8d44f6f9c97eb1038b831b}{\color{blue} countries-and-timezones} & 
    date \& timezones & 
    78 & 
    31 &
    115K & 
    7 & 
    7 (100.0\%) & 
    0 (0.0\%) & 
    7\\ 
\href{https://github.com/simple-statistics/simple-statistics/commit/31f037dd5550d554c4a96c3ee35b12e10a1c9cb7}{\color{blue} simple-statistics} & 
    statistics & 
    917 & 
    307 &
    103K & 
    89 & 
    3 (3.4\%) & 
    88 (98.9\%) & 
    3\\ 
\href{https://github.com/swang/plural/commit/f0027d66ecb37ce0108c8bcb4a6a448d1bf64047}{\color{blue} plural} & 
    text processing & 
    53 & 
    14 &
    18K & 
    4 & 
    3 (75.0\%) & 
    0 (0.0\%) & 
    3\\ 
\href{https://github.com/felixge/node-dirty/commit/d7fb4d4ecf0cce144efa21b674965631a7955e61}{\color{blue} dirty} & 
    key-value store & 
    89 & 
    24 &
    9.7K & 
    27 & 
    5 (18.5\%) & 
    0 (0.0\%) & 
    2\\ 
\href{https://github.com/rainder/node-geo-point/commit/c839d477ff7a48d1fc6574495cbbc6196161f494}{\color{blue} geo-point} & 
    geographical coordinates & 
    76 & 
    10 &
    1.1K & 
    19 & 
    10 (52.6\%) & 
    0 (0.0\%) & 
    11\\ 
\href{https://github.com/chakrit/node-uneval/commit/7578dc67090f650a171610a08ea529eba9d27438}{\color{blue} uneval} & 
    serialization & 
    31 & 
    3 &
    417 & 
    1 & 
    1 (100.0\%) & 
    0 (0.0\%) & 
    1\\ 
\midrule 
\href{https://gitlab.com/demsking/image-downloader/commit/19a53f652824bd0c612cc5bcd3a2eb173a16f938}{\color{blue} image-downloader} & 
    image handling & 
    32 & 
    12 &
    23K & 
    1 & 
    1 (100.0\%) & 
    0 (0.0\%) & 
    3\\ 
\href{https://gitlab.com/autokent/crawler-url-parser/commit/202c5b25ad693d284804261e2b3815fe66e0723e}{\color{blue} crawler-url-parser} & 
    URL parser & 
    100 & 
    185 &
    5 & 
    3 & 
    3 (100.0\%) & 
    0 (0.0\%) & 
    4\\ 
\href{https://gitlab.com/nerd-vision/opensource/gitlab-js/commit/c2c9ef54b1ea0fc82b284bc72dc2ff0935983f4c}{\color{blue} gitlab-js} & 
    API wrapper & 
    205 & 
    14 &
    184 & 
    37 & 
    4 (10.8\%) & 
    2 (5.4\%) & 
    7\\ 
\href{https://gitlab.com/cptpackrat/spacl-core/commit/fcb8511a0d01bdc206582cfacb3e2b01a0288f6a}{\color{blue} core} & 
    access control & 
    136 & 
    16 &
    1 & 
    20 & 
    6 (30.0\%) & 
    0 (0.0\%) & 
    2\\ 
\href{https://gitlab.com/comfort-stereo/omnitool/commit/0edf7d148337051c7c2307738423f0ff3db494c7}{\color{blue} omnitool} & 
    utility library & 
    1.6K & 
    420 &
    1 & 
    270 & 
    15 (5.6\%) & 
    80 (29.6\%) & 
    9\\ 
\bottomrule
  \end{tabular}
}
\vspace{-0.4cm}
\end{table*}

In response to this failure, the Prompt Refiner will create the prompt shown in
Figure~\ref{fig:RefinerExample}(b) from which a passing test is generated. In
this test, the expected value in the assertion has been updated to \code{5}, as
per the assertion error message.
 
Note that all these tests look quite natural and similar to tests that a human
developer might write, and they exercise typical usage scenarios (rather than
edge cases) of the functions under test.
\section{Research Questions \& Evaluation Setup}
\label{sec:evalsetup-rqs}

\subsection{Research Questions}
Our evaluation aims to answer the following research questions.

\begin{enumerate}[label=\textbf{RQ\arabic*},leftmargin=*]

\item \label{rq:cov} \textit{How much statement coverage and branch coverage do tests generated by \testpilot achieve?} Ideally, the generated tests would achieve high coverage to ensure that most of the API's functionality is exercised. Given that our goal is to generate complete unit test suites (as opposed to bug finding), we measure statement coverage for passing tests \textit{only}. We report coverage on both the package level and function level.

\item \label{rq:nessie-comparison} \textit{How does \testpilot's coverage compare to Nessie~\cite{DBLP:conf/icse/ArtecaHPT22}?} We compare \testpilot's coverage to the state-of-the-art JavaScript test generator, Nessie, which uses a feedback-directed approach.

\item \label{rq:non-trivial} \textit{How many of \testpilot's generated tests contain non-trivial assertions?} A test with no assertions or with \textit{trivial} assertions such as \code{assert.equal(true, true)} may still achieve high coverage. However, such tests do not provide useful oracles. We examine the generated tests and measure the prevalence of non-trivial assertions.

\item \label{rq:failing-tests} \textit{What are the characteristics of \testpilot's failing tests?} We investigate the reasons behind any failing generated test.

\item \label{rq:prompt-info} \textit{How does each of the different types of information included in prompts contribute to the effectiveness of \testpilot's generated tests?} To investigate if all the information included in prompts through the refiners is necessary to generate effective tests, we disable each refiner and report how it affects the results.

\item \label{rq:memorization} \textit{Are \testpilot's generated tests copied from existing tests?} Since \gptturbo is trained on GitHub code, it is likely that the LLM has already seen the tests of our evaluation packages before and may simply be producing copies of tests it ``memorized''. We investigate the similarity between the generated tests and any existing tests in our evaluation packages.

\item \label{rq:llm-comparison} \textit{How much does the coverage of \testpilot's generated tests rely on the underlying LLM?} To understand the generalizability of an LLM-based test generation approach and the effect of the underlying LLM \testpilot relies on, we compare coverage we obtain using \gptturbo with two other LLMs: 
(1) OpenAI's \cushman model \cite{CushmanSite}, one of \gptturbo's predecessors which is part of the Codex suite of LLMs \cite{CodexSite}  and which served as the main model behind the first release of GitHub Copilot \cite{CopilotSite}, 
and (2) \starcoder \cite{StarCoderSite}, a publicly available LLM for which the training process is fully documented. 
\end{enumerate}

\subsection{Evaluation Setup}

  \begin{table*}[t!]
  \centering
  \caption{Statement and branch coverage for \testpilot's passing tests, generated using \gptturbo. We also show passing tests that uniquely cover a statement. The last two columns show Nessie's statement and branch coverage for each package. Note that Nessie generates 1000 tests per package and the reported coverage is for all generated tests.}
  \label{tab:general-coverage}
  \resizebox{0.95\textwidth}{!}{
    \begin{tabular}{lrrrrrrrrr}
    \toprule
    \multirow{2}{*}{\textbf{Project}} & 
    \multicolumn{2}{c}{\textbf{Loading Coverage}} &
    \multicolumn{5}{c}{\textbf{\testpilot}} & 
    \multicolumn{2}{c}{\textbf{Nessie 1000 Tests}}\\
    \cmidrule(lr){2-3}\cmidrule(lr){4-8}\cmidrule(lr){9-10}
    & 
    \textbf{Stmt Cov} & \textbf{Branch Cov} &
    \textbf{Total Tests} & \textbf{Passing Tests (\%)} & \textbf{Stmt Cov} & \textbf{Branch Cov} & \textbf{Uniquely Contr. (\%)} &
    \textbf{ Stmt Cov }& \textbf{Branch Cov} \\ 
      \midrule

    glob 
    & 7.0\%
    & 0.4\%
    & 68 
    & 18 (26.5\%) 
    & \textbf{71.3\%} 
    & \textbf{66.3\%} 
    & 4 (22.2\%)
    & 39.7\%
    & 14.8\% \\ 

    fs-extra 
    & 16.8\%
    & 0.9\%
    & 471 
    & 277 (58.8\%) 
    & \textbf{58.8\%} 
    & \textbf{38.9\%} 
    & 17 (6.1\%)
    & 38.0\%
    & 24.9\% \\ 

    graceful-fs 
    & 28.6\%
    & 9.8\%
    & 345 
    & 177 (51.4\%) 
    & 49.3\% 
    & 33.3\% 
    & 1 (0.6\%)
    & \textbf{49.8\%}
    & \textbf{34.9\%} \\ 

    jsonfile 
    & 19.1\%
    & 0.0\%
    & 13 
    & 6 (48.0\%) 
    & 38.3\% 
    & 29.4\% 
    & 0 (0.0\%)
    & \textbf{91.5\%}
    & \textbf{81.0\%} \\ 

    bluebird 
    & 23.7\%
    & 7.8\%
    & 370 
    & 204 (55.2\%) 
    & \textbf{68.0\%} 
    & \textbf{50.0\%} 
    & 26 (12.5\%)
    & 43.8\%
    & 24.6\% \\ 

    q 
    & 22.4\%
    & 9.1\%
    & 323 
    & 186 (57.6\%) 
    & \textbf{70.4\%} 
    & 53.7\% 
    & 20 (10.5\%)
    & 66.8\%
    & \textbf{54.4\%} \\ 

    rsvp 
    & 16.4\%
    & 12.6\%
    & 109 
    & 70 (64.2\%) 
    & \textbf{70.1\%} 
    & \textbf{55.3\%} 
    & 6 (7.9\%)
    & 52.8\%
    & 47.0\% \\ 

    memfs 
    & 29.3\%
    & 7.2\%
    & 1037 
    & 471 (45.4\%) 
    & \textbf{81.1\%} 
    & \textbf{58.9\%} 
    & 40 (8.5\%)
    & 64.6\%
    & 36.2\% \\ 

    node-dir 
    & 5.9\%
    & 0.0\%
    & 40 
    & 19 (48.1\%) 
    & 64.3\% 
    & 50.8\% 
    & 4 (21.1\%)
    & \textbf{65.4\%}
    & \textbf{54.3\%} \\ 

    zip-a-folder 
    & 16.0\%
    & 0.0\%
    & 11 
    & 6 (54.5\%) 
    & 84.0\% 
    & 50.0\% 
    & 0 (0.0\%)
    & \textbf{88.0\%}
    & \textbf{100.0\%} \\ 
\midrule 

    js-sdsl 
    & 7.9\%
    & 3.7\%
    & 409 
    & 46 (11.3\%) 
    & \textbf{33.9\%} 
    & \textbf{24.3\%} 
    & 18 (39.1\%)
    & 8.5\%
    & 4.8\% \\ 

    quill-delta 
    & 8.1\%
    & 1.6\%
    & 152 
    & 33 (21.7\%) 
    & \textbf{73.0\%} 
    & \textbf{64.3\%} 
    & 8 (24.2\%)
    & 9.6\%
    & 2.5\% \\ 

    complex.js 
    & 8.4\%
    & 4.6\%
    & 209 
    & 121 (58.0\%) 
    & \textbf{70.2\%} 
    & \textbf{46.5\%} 
    & 10 (8.3\%)
    & 8.6\%
    & 5.4\% \\ 

    pull-stream 
    & 18.1\%
    & 0.0\%
    & 83 
    & 34 (41.0\%) 
    & \textbf{69.1\%} 
    & \textbf{52.8\%} 
    & 11 (32.4\%)
    & 38.5\%
    & 23.8\% \\ 

    countries-and-timezones 
    & 4.9\%
    & 0.0\%
    & 28 
    & 13 (46.4\%) 
    & 93.1\% 
    & 69.1\% 
    & 2 (15.4\%)
    & \textbf{96.0\%}
    & \textbf{80.8\%} \\ 

    simple-statistics 
    & 2.6\%
    & 0.0\%
    & 353 
    & 250 (70.9\%) 
    & \textbf{87.8\%} 
    & \textbf{71.3\%} 
    & 14 (5.4\%)
    & 57.8\%
    & 66.0\% \\ 

    plural 
    & 53.8\%
    & 0.0\%
    & 13 
    & 8 (61.5\%) 
    & \textbf{73.8\%} 
    & \textbf{59.1\%} 
    & 1 (12.5\%)
    & 59.2\%
    & 9.1\% \\ 

    dirty 
    & 4.7\%
    & 0.0\%
    & 70 
    & 32 (45.3\%) 
    & \textbf{74.5\%} 
    & \textbf{65.4\%} 
    & 2 (6.3\%)
    & 4.7\%
    & 0.0\% \\ 

    geo-point 
    & 12.2\%
    & 0.0\%
    & 76 
    & 50 (65.8\%) 
    & \textbf{87.8\%} 
    & \textbf{70.6\%} 
    & 1 (2.0\%)
    & 13.3\%
    & 0.0\% \\ 

    uneval 
    & 9.4\%
    & 0.0\%
    & 7 
    & 2 (28.6\%) 
    & 68.8\% 
    & 58.3\% 
    & 0 (0.0\%)
    & --
    & -- \\ 
\midrule 

    image-downloader 
    & 24.2\%
    & 0.0\%
    & 5 
    & 4 (80.0\%) 
    & \textbf{63.6\%} 
    & \textbf{50.0\%} 
    & 0 (0.0\%)
    & 30.3\%
    & 22.2\% \\ 

    crawler-url-parser 
    & 7.2\%
    & 1.3\%
    & 14 
    & 2 (14.3\%) 
    & 51.4\% 
    & 35.0\% 
    & 2 (100.0\%)
    & \textbf{73.9\%}
    & \textbf{64.1\%} \\ 

    gitlab-js 
    & 26.9\%
    & 0.6\%
    & 141 
    & 14 (9.9\%) 
    & 51.7\% 
    & 16.5\% 
    & 7 (46.4\%)
    & \textbf{55.3\%}
    & \textbf{26.4\%} \\ 

    core 
    & 16.1\%
    & 0.0\%
    & 85 
    & 13 (15.3\%) 
    & \textbf{78.3\%} 
    & \textbf{50.0\%} 
    & 5 (38.5\%)
    & 18.9\%
    & 0.0\% \\ 

    omnitool 
    & 19.2\%
    & 0.6\%
    & 1033 
    & 330 (31.9\%) 
    & \textbf{74.2\%} 
    & \textbf{55.2\%} 
    & 90 (27.2\%)
    & 56.0\%
    & 28.3\% \\ 
\midrule 
\textbf{Median} 
  & 16.1\%
  & 0.4\%
  &
  & 48.0\% 
  & \textbf{70.2\%} 
  & \textbf{52.8\%} 
  & 10.5\%  
  & 51.3\%
  & 25.6\% \\ 
\bottomrule
\end{tabular}
}
\vspace{-0.5cm}
\end{table*}

To answer the above research questions, we use a benchmark of \gptturbonumPackages npm packages.
Table~\ref{tab:packages} shows the size and number of downloads (popularity) of each of these packages. The first 10 packages shown in the table are the same GitHub-hosted packages used for evaluating Nessie~\cite{DBLP:conf/icse/ArtecaHPT22}, a recent feedback-directed test-generation technique for JavaScript.
However, we notice that these 10 packages primarily focus on popular I/O-related libraries with a callback-heavy style, so we add 10 new packages from different domains (e.g., document processing and data structures), programming styles (primarily object-oriented), as well as less popular packages.
Since \gptturbo (as well as the other LLMs we experiment with in \ref{rq:llm-comparison}) was trained on GitHub repositories, we have to assume that all our subject packages (and in particular their tests) were part of the model's training set.
For this reason, we also include an additional 5 packages whose source code is hosted on GitLab.%
\footnote{We checked similarly-named repos to ensure that they are not mirrored on GitHub.}

Table~\ref{tab:packages} shows that the \gptturbonumPackages packages vary in terms of popularity (downloads/week) and size (LOC), as well as in terms of the number of API functions they offer and the extent of the available documentation.
The ``API functions'' columns show the number of available API functions;
the number and proportion of API functions that have at least one example code snippet in the documentation (``w/ examples'');
and the number and proportion of API functions that have a documentation comment (``w/ comment'').
We also show the total number of example snippets available in the documentation of each package.

To answer \ref{rq:cov}--\ref{rq:memorization}, we run \testpilot using the \gptturbo LLM (version \textit{gpt-3.5-turbo-0301}), sampling five completions of up to 100 tokens at temperature zero,%
\footnote{Intuitively speaking, the sampling temperature controls the randomness of the generated completions, with lower temperatures meaning less non-determinism. Language models encode their
input and output using a vocabulary of tokens, with commonly occurring sequences of characters (such as \code{require}, but also contiguous runs of space characters) represented by a single token.}
with all other options at their default values.
In \ref{rq:llm-comparison}, we use the same settings for \cushman and \starcoder, except that the sampling temperature for the latter is 0.01 since it does not support a temperature of zero.

Note that LLM-based test generation does not have a test-generation budget per se since it is not an infinite process.
Instead, we ask the LLM for at most five completions for every prompt (but the model may return less).
We deduplicate the returned tests to avoid inflating the number of generated tests.
For example, if two prompts return the same test (modulo comments), we only record this test once but keep track of which prompt(s) resulted in its generation.

While we set the sampling temperature as low as possible, there is still some nondeterminism in the received responses.
Accordingly, we run all experiments 10 times.
All the per-package data points reported in Section~\ref{sec:eval} are median values over these 10 runs, except for integer-typed data such as number of tests where we use the ceiling of the median value.
For \ref{rq:memorization}, without loss of generality, we present the similarity numbers based on the first run only.

We use Istanbul/nyc~\cite{nyc} to measure statement and branch coverage and use Mocha's default time limit of 2s per test.

\section{Evaluation Results}
\label{sec:eval}

\subsection{\ref{rq:cov}: \testpilot's Coverage}

Table~\ref{tab:general-coverage} shows the number of tests \testpilot generates for each package, the number (and proportion) of passing tests, and the corresponding coverage achieved by the passing tests.
The first two columns of Table~\ref{tab:general-coverage} also show the coverage obtained by simply loading the package (\textit{loading coverage}).
This is the coverage we get ``for free'' without having any test suite, which we provide as a point of reference for interpreting our results.
Overall, \gptturbominPercentPassing --\gptturbomaxPercentPassing of the tests generated by \testpilot are passing tests, with a median of \gptturbomedianPercentPassing across all packages.
We now discuss the different coverage measurements of these passing tests.

\paragraph*{\textbf{Statement Coverage}} The statement coverage per package achieved by the passing tests ranges between \gptturbominstmtCoverage and \gptturbomaxstmtCoverage, with a median of \gptturbomedianstmtCoverage.
We note that across all packages, the achieved statement coverage is much higher than the loading coverage with a difference of \gptturbomindiffstmtCoverageWithLoading -- \gptturbomaxdiffstmtCoverageWithLoading and a median difference of \gptturbomediandiffstmtCoverageWithLoading.%
\footnote{For some of the projects we share with Nessie, our loading coverage differs from the one reported in their paper. We contacted the authors, who confirmed that with recent versions of Istanbul/nyc they obtained the same numbers as we did,
except for a very small difference on \texttt{memfs} (29.1\% vs 29.3\%), which may be due to platform differences.}

The lowest statement coverage \testpilot achieves is on \texttt{js-sdsl}, at \gptturbominstmtCoverage.
Upon further investigation of this package, we find that it maintains the documentation examples that appear on its website as markdown files in a separate repository.\footnote{https://github.com/js-sdsl/js-sdsl.github.io/tree/main/start} 
Including the extracted example snippets from this external repository increases the achieved coverage to 43.6\%, which suggests the importance of including usage examples in the prompts.
We examine the effect of the information included in prompts in detail in \ref{rq:prompt-info} (Section~\ref{sec:rq5-promptinfo}).

It is worth noting that \testpilot's coverage for the  GitLab projects listed in the bottom 5 rows of Table~\ref{tab:general-coverage} ranges from \gptturbominGitLabstmtCoverage to \gptturbomaxGitLabstmtCoverage.
This demonstrates that \testpilot is effective at generating high-coverage unit tests for packages it has not seen in its training set.

\paragraph*{\textbf{Branch Coverage}} We also show the branch coverage achieved by the passing tests in Table~\ref{tab:general-coverage}. We find that the branch coverage per package is between \gptturbominbranchCoverage and \gptturbomaxbranchCoverage, with a median of \gptturbomedianbranchCoverage. 
Similar to statement coverage, the achieved branch coverage is also much higher than the loading coverage with a difference of \gptturbomindiffbranchCoverageWithLoading -- \gptturbomaxdiffbranchCoverageWithLoading and a median difference of \gptturbomediandiffbranchCoverageWithLoading.

Since achieving branch coverage is generally harder than achieving statement coverage, it is expected that the branch coverage for the generated tests
is lower than the statement coverage.
However, we note an interesting case in \texttt{gitlab-js} where this difference seems more pronounced (51.7\% vs. 16.5\%). 
Upon further investigation of its source code and documentation, we find that \texttt{gitlab-js} offers various configuration options and parameters to specify the GitLab repository to connect to and use/query (e.g., its url, authentication token, search parameters to use for a query).
The processing of these options is reflected in the main branching logic in the code.
While \testpilot does attempt to generate reasonable tests that call different endpoints with different options, it sometimes struggles to find the correct function call to use, resulting in type errors.
In general, a large proportion of the tests \testpilot generates for this package fail, and thus do not contribute to our coverage numbers.
It is also worth noting that properly testing such a package would require mocking, but we did not observe any of the generated tests to use mocking.
In the future, it would be interesting to investigate if including mocking libraries in the prompt, or other mocking related information, may result in the model using mocking when needed.

\begin{figure}[t!]
\centering
\includegraphics[width=0.48\textwidth]{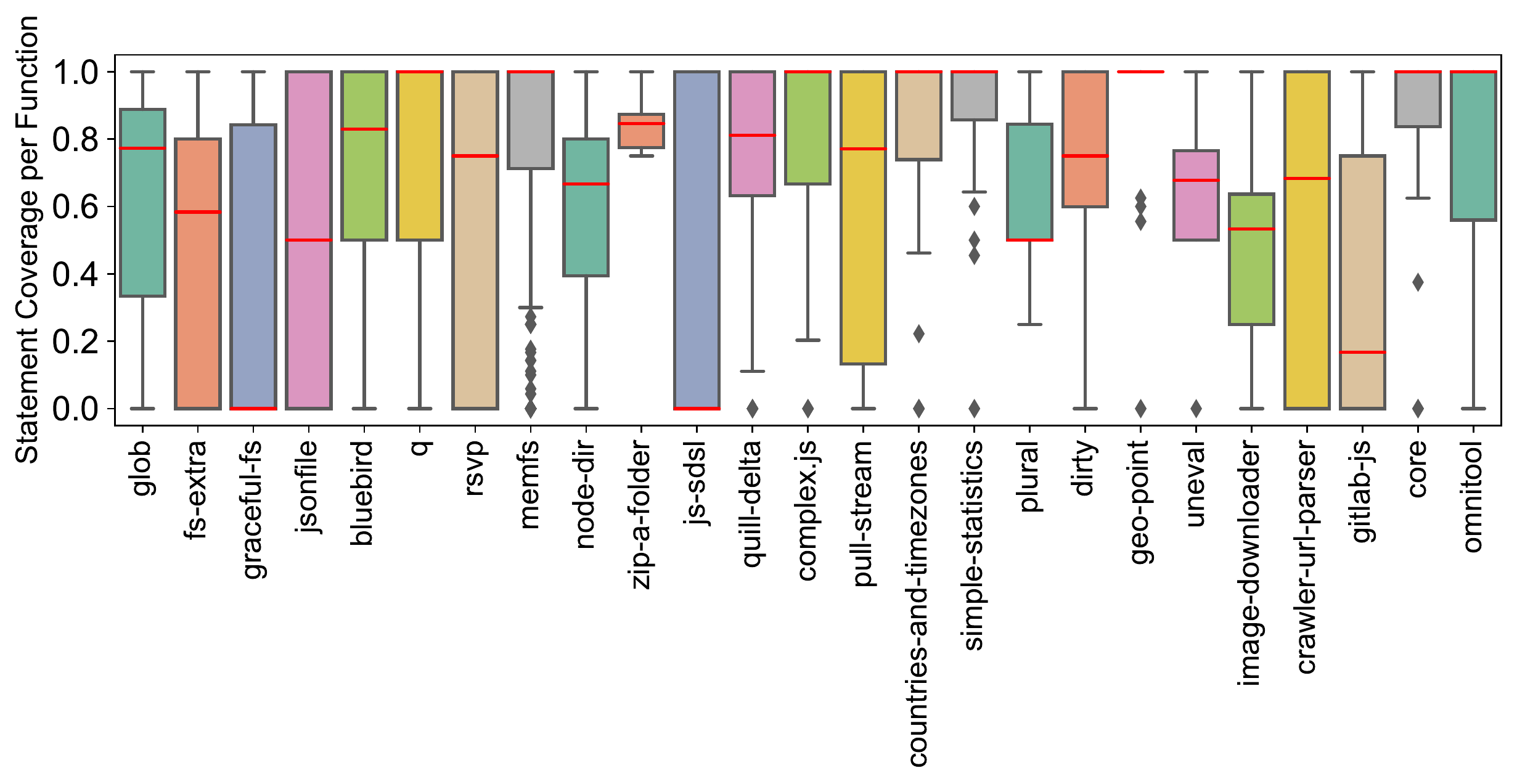}
\caption{Distribution of statement coverage per function for \testpilot's generated tests using \gptturbo.}
\label{fig:function-coverage}
\end{figure}

\paragraph*{\textbf{Coverage per function}} Figure~\ref{fig:function-coverage} shows the distribution of statement coverage per function for each package.
Each box corresponds to one of our benchmark packages and each data point in a box represents the statement coverage for a function in that package.
The median statement coverage per function for each package is shown in red.

Overall, the median statement coverage per function for a given project ranges from \gptturbominMedianFnCovPerProj--\gptturbomaxMedianFnCovPerProj, with a median of \gptturbomedianMedianFnCovPerProj. 
To ensure that \testpilot is not generating high coverage tests only for smaller functions, we run a Pearson's correlation test between the statement coverage per function and the corresponding function size (in statements).
We find no statistically significant correlation between coverage and size, indicating that \testpilot is not only doing well for smaller functions.\footnote{Exact correlation coefficients and p-values are provided in our artifact.}

As expected, Figure~\ref{fig:function-coverage} shows that for most packages, \testpilot does well for some functions while achieving low coverage for others.
Let us take \texttt{jsonfile} as an example.
In Table~\ref{tab:general-coverage}, we saw that its statement coverage at the package level is 38.3\%.
From Figure~\ref{fig:function-coverage}, we see that statement coverage per function ranges from 0\% to 100\%, with a median of almost 50\%.
Diving into the data, we find that there are two functions that \testpilot cannot cover, because their corresponding generated tests fail either due to references to non-existent files \testpilot includes in the tests or because they time out.
However, the functions that \testpilot is able to cover have statement coverage ranging from 58\%-100\%.
We can observe similar behavior with other file system dependent packages, such as \texttt{graceful-fs} or \texttt{fs-extra}.
At the other end of the spectrum, we see \texttt{zip-a-folder} where \testpilot achieves both high statement coverage at the package level (84\%) as well as high statement coverage at the function level in Figure~\ref{fig:function-coverage} where the minimum per function coverage is 75\%. 
 
\paragraph*{\textbf{Uniquely Contributing Tests}} To further understand the diversity of the generated tests, Table~\ref{tab:general-coverage} also shows how many of the tests \testpilot generates are
\textit{uniquely contributing}, meaning that they cover at least one statement that no other tests cover.
A median of \gptturbomedianPercentUniquelyCoveringTests of the passing tests are of this kind, with some packages as high as \gptturbomaxPercentUniquelyCoveringTests.
These results are promising because they show that \testpilot can generate tests that cover edge cases, but there is clearly some redundancy among the generated tests. 
Of course, we cannot simply exclude all \gptturboremainingNonUniqueTests remaining tests without losing coverage, since some statements may be covered by multiple tests non-uniquely.
Exploring test suite minimization techniques~\cite{yoo2012regression} to reduce the size of the generated test suite is an interesting avenue for future work.

\subsection{\ref{rq:nessie-comparison} \testpilot vs. Nessie~} \label{sec:NessieComparison}
We compare \testpilot's coverage to the state-of-the-art JavaScript test generator Nessie~\cite{DBLP:conf/icse/ArtecaHPT22}, which uses a traditional feedback-directed approach.%
\footnote{Note that Nessie's implementation has been refactored and improved after the publication of the original paper, which is why some of the values in this table differ slightly from the published numbers.
Nessie's first author has kindly helped us run the improved version (specifically, \url{https://github.com/emarteca/nessie/tree/86e48f1d038d98dcd2663d6d36a58a70c4666b1b}) on all 25 packages. We include the Nessie results in our artifact}
For each package, Nessie generates 1000 tests, for which we measure statement and branch coverage in the same way as for \testpilot.
We then repeat these measurements 10 times and take the median coverage across the 10 runs to follow a similar setup to \testpilot's evaluation.
We use a Wilcoxon paired rank-sum test to determine if there are statistically significant differences between the coverage achieved by both tools.

\begin{figure*}
\centering
\begin{tabular}{c}
  \begin{minipage}[t]{0.95\textwidth}
    \begin{minted}[linenos,mathescape,escapeinside=||,fontsize=\footnotesize,highlightlines={4,6-9,11,18-21}]{javascript}
let manuelmhtr_countries_and_timezones = require("../../TEST_REPO_manuelmhtr_countries_and_timezones");
module.exports = function() {
	let ret_val_manuelmhtr_countries_and_timezones_1;
	try { 
		ret_val_manuelmhtr_countries_and_timezones_1 = manuelmhtr_countries_and_timezones.getAllCountries(); |\label{line:NessieTest:getAllCountries}|
		Promise.resolve(ret_val_manuelmhtr_countries_and_timezones_1).catch(e => { console.log({"error_1": true}); });
	} catch(e) {
		console.log({"error_1": true});
	}
	let ret_val_manuelmhtr_countries_and_timezones_2;
	try { 
		ret_val_manuelmhtr_countries_and_timezones_2 = 
		  manuelmhtr_countries_and_timezones.getCountry({"k": -293.76984807333383, |\label{line:NessieTest:getCountryStart}|
		                                                 "rHMR": -17.71151399309167, 
		                                                 "vSF6": 721.0602634375625, 
		                                                 "l": 497.17371230897766, 
		                                                 "EnL": -611.9090030925536});|\label{line:NessieTest:getCountryEnd}|
		Promise.resolve(ret_val_manuelmhtr_countries_and_timezones_2).catch(e => { console.log({"error_2": true}); });
	} catch(e) {
		console.log({"error_2": true});
	}
}

    \end{minted}
  \end{minipage}
\end{tabular}
\caption{Example of a test generated by Nessie. Highlighted lines are for debugging purposes only and do not contribute to the testing of the package under test.}
  \label{fig:ExampleNessieTest}
\end{figure*}

The last two columns of Table~\ref{tab:general-coverage} show statement and branch coverage for Nessie.
We note that Nessie could not run on \texttt{uneval}, because the module's only export is a function, which Nessie does not support.
For the remaining 24 packages, Nessie achieved \gptturbominNessiestmtCoverage -- \gptturbomaxNessiestmtCoverage statement coverage, with a median of \gptturbomedianNessiestmtCoverage.
In contrast, as shown in Table~\ref{tab:general-coverage}, \testpilot's median statement coverage is much higher at \gptturbomedianstmtCoverage.
The difference in branch coverage is even higher, with \gptturbomedianbranchCoverage for \testpilot vs \gptturbomedianNessiebranchCoverage for Nessie.
Both these differences are statistically significant (p-values 0.002 and 0.027 respectively) with a large effect size, measured by Cliff's delta~\cite{cliff1993dominance}, of 0.493 for statement coverage and a medium one (0.431) for branch coverage.\footnote{All effect sizes for all statistical tests are available in our artifact.}
Note that Nessie always generates 1000 tests per package, while \testpilot usually generates far fewer tests, except on \texttt{memfs} and \texttt{omnitool}.
It is also worth emphasizing that Nessie (and other test-generation techniques such as LambdaTester~\cite{lambdatesterPaper}) report coverage of \textit{all} generated tests, regardless of whether they pass or fail while our reported coverage numbers are for passing tests only.

We now dive into the results at the package level.
For each package, Table~\ref{tab:general-coverage} highlights the higher coverage from the two techniques in bold.
\testpilot outperforms Nessie on \gptturbonumProjTestPilotHigherThanNessiestmtCoverage of the 24 packages (\gptturboprojsTestPilotHigherThanNessiestmtCoverage), increasing coverage by \gptturbomintpVsNessieHigherstmtCoverageDiff --\gptturbomaxtpVsNessieHigherstmtCoverageDiff, with a median \gptturbomediantpVsNessieHigherstmtCoverageDiff increase.
For \gptturbonumProjTestPilotLowerThanNessiestmtCoverage of the remaining packages (\gptturboprojsTestPilotLowerThanNessiestmtCoverage), \testpilot achieves lower coverage than Nessie.
For these packages, it reduces coverage by \gptturbomintpVsNessieLowerstmtCoverageDiff -- 
\gptturbomaxtpVsNessieLowerstmtCoverageDiff, with a median \gptturbomediantpVsNessieLowerstmtCoverageDiff decrease. 
We also note that Nessie fails to achieve any branch coverage on \gptturbonumNessieZeroBranchCov projects (\gptturboNessieZeroBranchCovProjs), while the statement coverage for these projects is non-zero.
Upon further examination, and after consulting the Nessie authors, we found that Nessie cannot generate tests that instantiate classes,
meaning that statement coverage is barely above loading coverage for packages with a class-based API, while the branch coverage is zero.

Aside from the difference in coverage achieved by Nessie and \testpilot, tests generated by Nessie tend to look quite different
from the ones generated by \testpilot, which stems from Nessie's random approach to test generation. To illustrate this, Figure~\ref{fig:ExampleNessieTest}
shows an example of a test generated by Nessie that exercises the \code{getCountry} function of \texttt{countries-and-timezones}.
As can be seen in the figure, the test uses long variable names such as \code{ret_val_manuelmhtr_countries_and_timezones_1} that hamper
readability. Moreover, the test invokes \code{getCountry} on lines~\ref{line:NessieTest:getCountryStart}--\ref{line:NessieTest:getCountryEnd} with 
an object literal that binds random values to some randomly named properties, which does not reflect intended use of the API. 
Moreover, tests generated by Nessie do not contain any assertions.  By contrast, tests generated by \testpilot for the same package
(see Figure~\ref{fig:PromptExamples}) typically use variable names that are similar to those chosen by programmers, invoke APIs with sensible
values, and often contain assertions.

\subsection{\ref{rq:non-trivial}: Non-trivial Assertions}

We define a \textit{non-trivial assertion} as an assertion that depends on at least one function from the package under test. To identify non-trivial assertions, we first use CodeQL~\cite{codeql} to compute a backwards program slice from each assertion in the generated tests.
We consider assertions whose backwards slice contains an import of the package under test as non-trivial assertions. We then report generated tests that contain at least one non-trivial assertion.
 
Table~\ref{tab:nontrivial-coverage} shows the number of tests with non-trivial assertions (\textit{non-trivial test} for short) and their proportion w.r.t all generated tests from Table~\ref{tab:general-coverage}. The table also shows the number and proportion of these tests that pass, along with the statement coverage they achieve.

We observe that there is only one package, \texttt{image-downloader} where \testpilot generates only trivial tests.
While the generated tests for \texttt{image-downloader} did include calls to its API, they were all missing assert statements.
Across the remaining packages, a median of \gptturbosecondMinPercentNonTrivialTests~-- \gptturbomaxPercentNonTrivialTests of \testpilot's generated tests per package are non-trivial.
A median of \gptturbomedianPercentNonTrivialTests of the generated tests for a given package are non-trivial.
When compared to all generated tests, we can also see that only a slightly lower proportion of non-trivial tests pass (median \gptturbomedianPercentPassing for overall passing tests from Table~\ref{tab:general-coverage} vs. \gptturbomedianPercentNonTrivialTestsPassing for non-trivial passing tests from Table~\ref{tab:nontrivial-coverage}).
Both these results show that \testpilot typically generates tests with assertions that exercise functionality from the target package.

  \begin{table}[t!]
  \centering
  \caption{Number (\%) of \textbf{\textit{non-trivial}} \testpilot tests generated using \gptturbo and the resulting statement coverage from the passing non-trivial tests.}
  \label{tab:nontrivial-coverage}
  \resizebox{0.9\columnwidth}{!}{
  \begin{tabular}{lrrr}
  \toprule
  \multirow{2}{*}{\textbf{Project}} & \multirow{2}{*}{\textbf{\thead{Non-trivial \\Tests (\%)}}} & 
  \multicolumn{2}{c}{\textbf{Passing Non-trivial Tests}} \\
  \cmidrule{3-4}
  & & \textbf{Tests (\%)} & \textbf{Stmt Cov} \\ 
      \midrule
glob & 37 (54.4\%) & 3 (8.1\%) & 50.1\% \\ 
fs-extra & 142 (30.1\%) & 70 (49.5\%) & 28.0\% \\ 
graceful-fs & 64 (18.4\%) & 27 (42.5\%) & 41.5\% \\ 
jsonfile & 4 (32.0\%) & 0 (0.0\%) & 0.0\% \\ 
bluebird & 227 (61.4\%) & 137 (60.4\%) & 61.6\% \\ 
q & 235 (72.6\%) & 136 (58.0\%) & 66.4\% \\ 
rsvp & 68 (62.4\%) & 48 (70.6\%) & 67.6\% \\ 
memfs & 758 (73.1\%) & 356 (47.0\%) & 77.4\% \\ 
node-dir & 7 (16.5\%) & 0 (0.0\%) & 0.0\% \\ 
zip-a-folder & 1 (9.1\%) & 0 (0.0\%) & 0.0\% \\ 
\midrule 
js-sdsl & 349 (85.3\%) & 44 (12.6\%) & 33.9\% \\ 
quill-delta & 92 (60.5\%) & 27 (28.8\%) & 59.7\% \\ 
complex.js & 190 (90.9\%) & 104 (54.6\%) & 62.7\% \\ 
pull-stream & 60 (72.3\%) & 29 (47.5\%) & 64.7\% \\ 
countries-and-timezones & 22 (78.6\%) & 7 (31.8\%) & 73.5\% \\ 
simple-statistics & 189 (53.6\%) & 115 (60.6\%) & 46.9\% \\ 
plural & 12 (92.3\%) & 8 (66.7\%) & 73.8\% \\ 
dirty & 29 (41.7\%) & 13 (44.8\%) & 66.0\% \\ 
geo-point & 60 (78.9\%) & 34 (56.7\%) & 64.6\% \\ 
uneval & 4 (57.1\%) & 2 (50.0\%) & 68.8\% \\ 
\midrule 
image-downloader & 0 (0.0\%) & -- & 0.0\% \\ 
crawler-url-parser & 6 (42.9\%) & 1 (16.7\%) & 49.5\% \\ 
gitlab-js & 104 (73.8\%) & 12 (11.5\%) & 49.3\% \\ 
core & 64 (74.7\%) & 12 (18.9\%) & 75.5\% \\ 
omnitool & 977 (94.6\%) & 319 (32.6\%) & 73.8\% \\ 
\midrule 
\textbf{Median} & 61.4\% & 43.7\% & 61.6\% \\ 
\bottomrule
\end{tabular}
}
\vspace{-0.6cm}
\end{table}

The coverage achieved by the non-trivial tests also supports this finding.
Specifically, when comparing the statement coverage for all the generated tests in Table~\ref{tab:general-coverage} to that for non-trivial tests in Table~\ref{tab:nontrivial-coverage}, we find that
the difference ranges from \gptturbominDiffAllVsNonTrivial--\gptturbomaxDiffAllVsNonTrivial, with a median difference of only \gptturbomedianDiffAllVsNonTrivial.
This means that the achieved coverage for most packages mainly comes from exercising API functionality that is tested by the generated oracles.
We note however that there are \gptturbonumPkgsWithZeroNonTrivialCoverage packages (\gptturbopkgsWithZeroNonTrivialCoverage) where non-trivial tests achieve 0\% statement coverage, causing the larger differences.
Apart from \texttt{image-downloader} discussed above, the three remaining packages do not have any passing non-trivial tests.
Since we calculate coverage for passing tests only, this results in the 0\% statement coverage for the non-trivial tests.

\subsection{\ref{rq:failing-tests}: Characteristics of Failing Tests}

\begin{figure}[t!]
  \centering
  \includegraphics[width=0.5\textwidth]{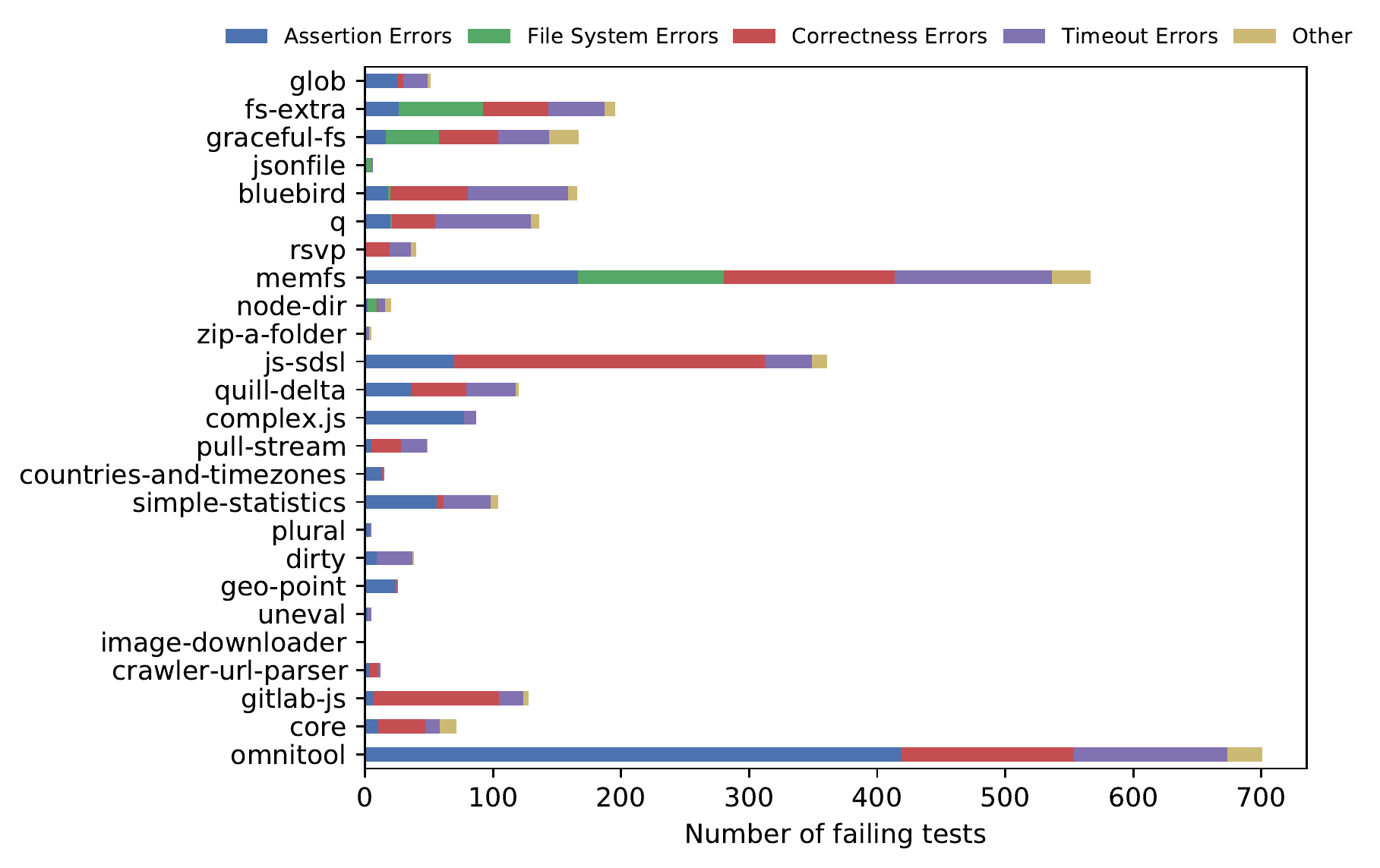}  
  \caption{Types of errors in the failed tests generated by \testpilot, using \gptturbo.}
    \label{fig:failure-reasons}
\end{figure}

Figure~\ref{fig:failure-reasons} shows the number of failing tests for each package, along with the breakdown of the reasons behind the failure.
Assertion errors occur when the expected value in an assertion does not match the actual value from executing the code.
File-system errors include errors such as files or directories not being found, which we identify by checking for
file-system related error codes~\cite{nodejsFSErrors} in the error stack trace.
Correctness errors include all type errors, syntax errors, reference errors, incorrect invocations of \texttt{done}, and infinite recursion/call stack errors.
Timeout errors occur when tests exceed the maximum running time we allow them (2s/test).
Finally, we group all other application-specific errors we observe under Other.

We find that the most common failure reason is timeouts with a median \gptturbomedianPercentTimeoutErrors of failing tests, followed by correctness errors (particularly type errors) with a median of \gptturbomedianPercentCorrectnessErrors of failing tests.
The majority of timeouts are due to missing calls to \texttt{done}, causing Mocha to keep waiting for the call.
We note that on average, the \RetryWithError refiner was able to fix \gptturboTimeoutErrorFixedWithRetry of such timeout errors, with the model often simply adding a call to \texttt{done}%
\footnote{
  While the insertion of missing calls to \texttt{done} may seem straightforward and therefore be amenable to automated repair, 
  it can be surprisingly tricky to find the correct locations where to insert such calls, and handling this correctly would 
  require applying static analysis to the generated test. We therefore opted for an automated approach that relies solely on the LLM
  but will consider the use of static analysis to repair generated tests as future work.
}.

We find that a median \gptturbomedianPercentAssertionErrors of failures are assertion errors, indicating that in some cases \gptturbo is not able to figure out the correct expected value for the test oracle.
This is especially true when the package under test is not widely used and none of the information we provide the model can help it in figuring out the correct values.
For example, in one of the tests for \texttt{geo-point}, \testpilot was able to use coordinates in the provided example snippet to correctly construct two geographical coordinates as input for the \texttt{calculateDistance} function, which computes the distance between the two coordinates.
However, \testpilot incorrectly generated 131.4158102876726 as the expected value for the distance between these two points, while the correct expected value is 130584.05017990958; this caused the test to fail with an assertion error.
We note that in this specific case, when \testpilot re-prompted the model with the failing test and error message, it was then able to produce a passing test with the corrected oracle. 
On average across the packages, we find that the \RetryWithError refiner was able to fix \gptturboAssertionErrorFixedWithRetry of assertion errors.

Finally, we note that file-system errors are domain specific.
The generated tests for packages in the file system domain, such as \texttt{fs-extra} or \texttt{memfs}, have a high proportion of failing tests due to such errors.
This is not surprising given that these tests may rely on files that may be non-existent or require containing specific content.
Packages in the other domains do not face this problem.

Overall, we find that re-prompting the model with the error message of failing tests (regardless of the failure reason) allows \testpilot to produce a consequent passing test in \gptturboallErrorsFixedWithRetry of the cases. 

\begin{figure}[t!]
\centering
\includegraphics[width=0.49\textwidth]{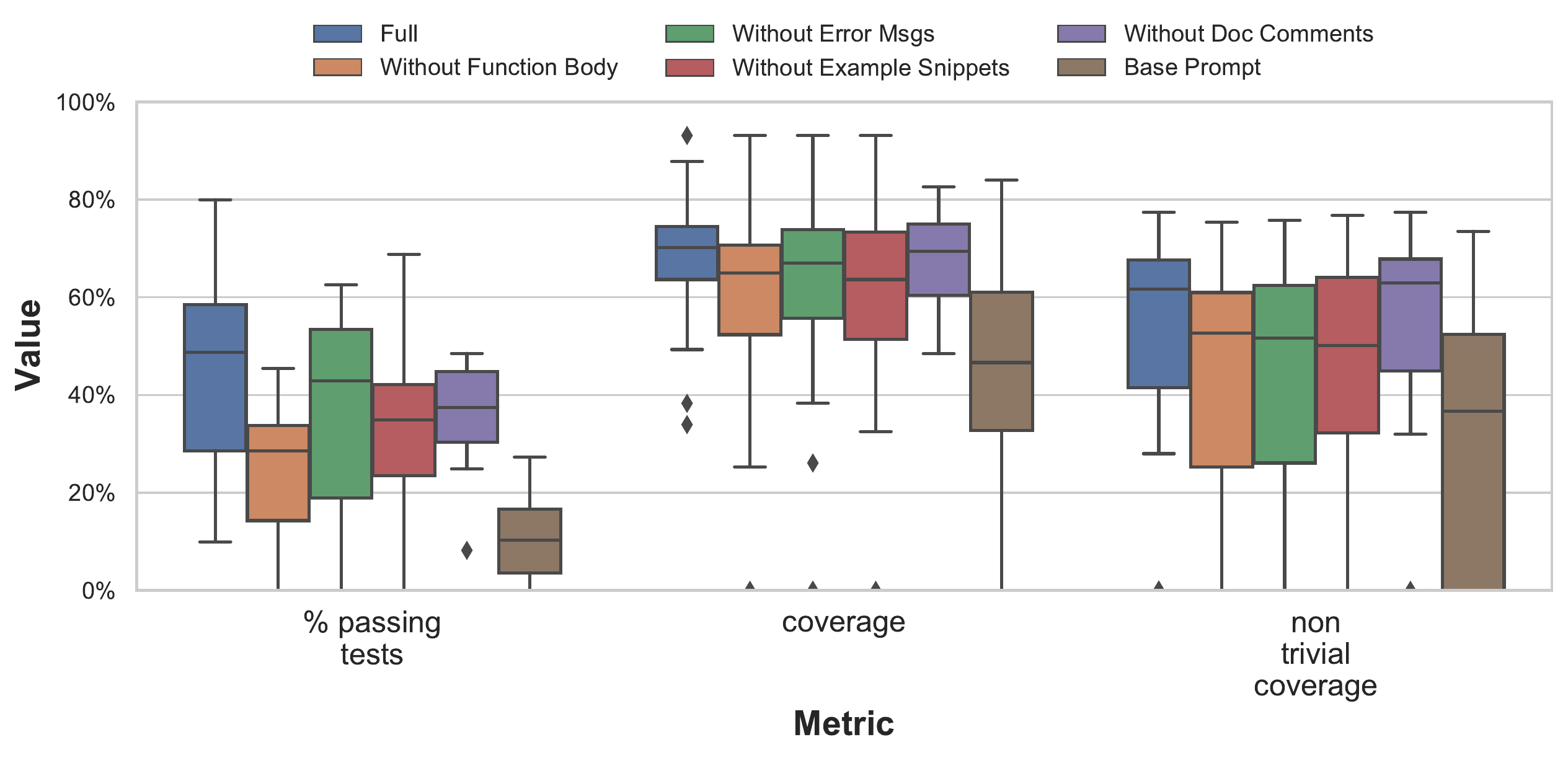}
\caption{Effect of disabling prompt refiners in \testpilot, using \gptturbo. \textit{Full} considers all refiners while \textit{Base} includes only the function signature and Mocha scaffolding.\vspace{-0.5cm}}
\label{fig:refiner-boxplot}
\end{figure}

\begin{figure*}[t!]
\centering
\begin{tabular}{c@{\hspace*{0.05\textwidth}}c}
  \begin{minipage}[t]{0.45\textwidth}
    \begin{minted}[linenos,mathescape,fontsize=\footnotesize,highlightlines={1-6}]{javascript}
let mocha = require('mocha');
let assert = require('assert');
let complex_js = require('complex.js');
// complex.js.ZERO.valueOf()
describe('test suite', function() {
    it('test case', function(done) {
        assert.equal(complex_js.ZERO.valueOf(), 0);
        done();
    })
})    
    \end{minted}
  \end{minipage}
&
  \begin{minipage}[t]{0.45\textwidth}
    \begin{minted}[linenos,fontsize=\footnotesize,highlightlines={1-12}]{javascript}
let mocha = require('mocha');
let assert = require('assert');
let complex_js = require('complex.js');
// complex.js.ZERO.valueOf()
// function() {
//   if (this['im'] === 0) {
//     return this['re'];
//   }
//   return null;
// }
describe('test complex_js', function() {
    it('test complex.js.ZERO.valueOf', function(done) {
      assert.equal(complex_js.ZERO.valueOf(), null);
      done();
    })
})
    \end{minted}
  \end{minipage}
  \\
  {\bf (a)} & {\bf (b)}
\end{tabular}
\caption{Example of a refinement negatively influencing test generation. 
  Prompt~(a) contains no information except the method signature, and the generated test passes.
  Prompt~(b) adds the body of the method, but the generated test fails.
}
  \label{fig:UselessRefinement}
\end{figure*}

\subsection{\ref{rq:prompt-info}: Effect of Prompt Refiners}
\label{sec:rq5-promptinfo}

Our results so far include tests generated with all four prompt refiners discussed in Section~\ref{sec:approach}.
In this RQ, we investigate the effect of each of these refiners on the quality of the generated tests.
Specifically, we conduct an ablation study where we disable one refiner at a time.
Disabling a refiner means that we no longer generate prompts that include the information it provides.
For example, disabling \DocCommentIncluder means that none of the prompts we generate would contain documentation comments.
We then compare the percentage of passing tests, the achieved coverage, as well as the coverage by non-trivial tests (\textit{non-trivial coverage}).

Figure~\ref{fig:refiner-boxplot} shows our results.
The x-axis shows the metrics we compare across the different configurations shown in the legend.
The y-axis shows the values for each metric (all percentages).
Each data point in a boxplot represents the results of the specific metric for a given package, using the corresponding refiner configuration.
The black line in the middle of each box represents the median value for each metric across all packages.
The full configuration is the configuration we presented so far (i.e., all refiners enabled).
The other configurations show the results of excluding only one of the refiners.
For example, the red box plot shows the results when disabling the \SnippetIncluder (i.e., \textit{Without Example Snippets}).
The base prompt configuration contains only the function signature and test scaffolding (i.e., disabling all refiners).
Note, however, that only \gptturbonumPackagesWithDocComments of the packages in our evaluation contain documentation comments.
It does not make sense to compare the effect of disabling the \DocCommentIncluder on packages that do not contain doc comments in the first place.
Therefore, while the distributions shown in all boxplots represent \gptturbonumPackages packages, the \textit{Without Doc Comments} configuration contains data for only \gptturbonumPackagesWithDocComments packages.

Overall, we can see that the full configuration outperforms all other configurations, across all three metrics, implying that all the prompt information we include contributes to generating more effective tests.
We find that there was not a single package where disabling a refiner led to better results on any metric.
With the exception of \gptturbonumPkgsWithUselessRefiner packages where disabling one of the refiners did not affect the results (\SnippetIncluder on \texttt{crawler-url-parser} and \texttt{dirty}; and \RetryWithError on \texttt{gitlab-js} and \texttt{zip-a-folder}), disabling a refiner \textit{always} resulted in lower values in at least one metric.

The contributions of the refiners are especially notable for the percentage of passing tests where disabling any of the refiners (e.g., \FunctionBodyIncluder or \SnippetIncluder) results in a large drop in the percentage of passing tests.
This suggests that the refiners are effective in guiding the model towards generating more passing tests, even if this does not necessarily result in additional coverage.
We find that across \textit{all} packages, a full configuration always leads to a higher percentage of passing tests for a given API, while maintaining high coverage.

To understand if the differences between the distributions we observe in Figure~\ref{fig:refiner-boxplot} are statistically significant, we compare the results of each pair of configurations for all three metrics using a Wilcoxon matched pairs signed rank tests.
Note that when comparing against \DocCommentIncluder, we compare distributions for only the \gptturbonumPackagesWithDocComments packages that contain doc comments.

We find a statistically significant difference between the full configuration and each configuration that disables \textit{any} refiner as well as between the base configuration and each of the other configurations.
Compared to the full configuration, the largest effect size we observed for disabling a refiner was on passing tests when either \FunctionBodyIncluder or \DocCommentIncluder were disabled (Cliff's delta 0.582 and 0.531 respectively).

Apart from differences with the full and base configuration, we find no statistically significant differences between the pairs of other configurations except for the following cases:
We find that for both passing tests and coverage, there is a statistically significant difference between the configuration that disables \FunctionBodyIncluder and that which disables \RetryWithError (medium and negligible effect sizes, respectively).
For passing tests, we also find a statistically significant difference between disabling \FunctionBodyIncluder and disabling each of \SnippetIncluder and \DocCommentIncluder (small and medium effect sizes, respectively).
However, we note that a sample size of \gptturbonumPackagesWithDocComments is too small to draw any valid conclusions for \DocCommentIncluder.
It is particularly interesting to see that there was no statistically significant difference between disabling \SnippetIncluder and disabling any of the other refiners.
This suggests that the absence of example snippets does not necessarily affect the metrics any more than the absence of any of the information provided by the other refiners.
Since Figure~\ref{fig:refiner-boxplot} shows that we still obtain a high median coverage even when disabling \SnippetIncluder, 
this suggests that the presence of examples snippets is not essential for generating effective test suites with high coverage, and that \testpilot is applicable even in cases where no documentation examples are present.

Finally, we note that while the overall results across a given package show that the refiners always improve, or at least maintain, coverage and percentage of passing tests, this does not mean that a refiner always improves the results for an individual API function. 
We have observed situations where adding information such as the function implementation to a prompt that does not
include it confuses the model, resulting in the generation of a failing test. Figure~\ref{fig:UselessRefinement} shows an example for the
\texttt{complex.js} package: given the base prompt on the left, \gptturbo is able to produce a (very simple) passing test for the \code{valueOf}
method of the constant \code{ZERO} exported by the package; adding the function body yields the prompt on the right, which seems to confuse
the model, resulting in the generation of a failing test. Across all packages, 5,367 prompts were generated by applying one of the refiners,
and in only 394 cases (7.3\%) the refined prompt was less effective than the original prompt in the sense that a passing test was generated from
the original prompt, but not from the refined prompt.

\subsection{\ref{rq:memorization}: Memorization}

\begin{figure}[t!]
\centering
\includegraphics[width=0.5\textwidth]{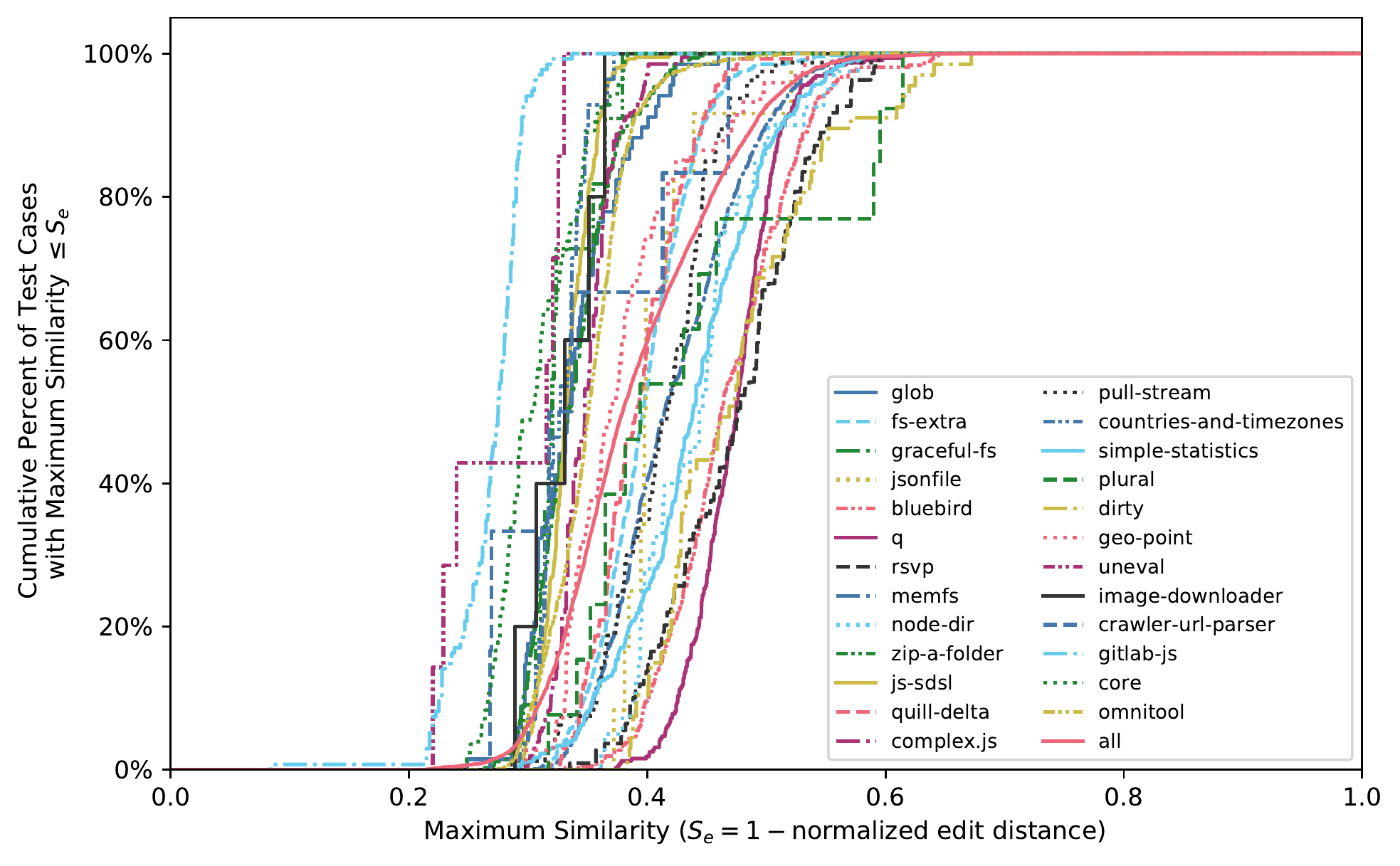}
\caption{Cumulative percent of \testpilot generated test cases, using \gptturbo, with maximum similarity less than the similarity value shown on the x-axis.}
\label{fig:cumulative_sim}
\end{figure}

Since \gptturbo was trained on GitHub code, some of the existing tests included in our benchmarks may have been part of its training set.
This raises the concern that \testpilot may be memorizing existing tests, rather than generating new ones, limiting its usefulness for packages it was not trained on.
To investigate potential effects of memorization, we measure the similarity between each generated test and the existing tests in the benchmarks (number of existing tests shown in Table~\ref{tab:packages}).
Recently, Lemieux et al.~\cite{LemieuxICSE2023} reported that code plagiarism or clone detection~\cite{SchleimerPlagiarism} techniques are not effective at identifying LLM code memorization.
Instead, they find that measuring similarity through edit distance~\cite{MyersEditDistance} produces more meaningful results. 
They define \textit{maximum similarity} as a metric that measures the normalized highest similarity between a given generated test and all existing tests as follows:
$\max_{tp \in \mathit{TP}} \left( 1 - \frac{\mathit{dist}(t^*, t_p)}{\max(\mathit{len}(t^*), \mathit{len}(tp))}\right)$, where $\mathit{TP}$ is the set of existing test functions in a package,
$t^*$ is a given generated test, and $\mathit{dist}$ is the edit distance between a generated test and an existing test.
We follow the same method for calculating maximum similarity for each generated test, using the npm Levenstein package~\cite{levenstein} to calculate $\mathit{dist}$.

\begin{figure}[t!]
\centering
\begin{tabular}{c}
  \begin{minipage}[t]{0.4\textwidth}
    \begin{minted}[mathescape,fontsize=\footnotesize]{javascript}
it('test case', function(done) {
    bluebird.resolve().then(function() {
        throw new Error('test');
    }).catchThrow().catch(function(err) {
        assert.equal(err.message, 'test');
    }).finally(done);
});
\end{minted}
  \end{minipage}
  \\
{\bf (a)}\\
\\
  \begin{minipage}[t]{0.4\textwidth}
    \begin{minted}[fontsize=\footnotesize]{javascript}
it("1 level", function() {
    return Promise.resolve().then(function() {
        throw new Error();
    }).then(assert.fail, function(e) {
        assertLongTrace(e, 1 + 1, [1]);
    });
});
  \end{minted}
  \end{minipage}
  \\
  {\bf (b)}
\end{tabular}
\caption{ 
  Example of a \testpilot-generated test case for \texttt{bluebird} (a), and an existing test case (b) with similarity 0.62.\vspace{-0.4cm}
}
  \label{fig:highSimilarity}
\end{figure}

\begin{table*}[t!]
\centering
\caption{A comparison of statement coverage of \testpilot's generated tests using three LLMs. For each project, we show the number of generated tests, the number (\%) of passing tests, and the statement coverage achieved by these passing tests.}
\label{tab:llm-comparison}
\resizebox{\textwidth}{!}{
\begin{tabular}{lrrrrrrrrrrrr}
\toprule
\multirow{2}{*}{\textbf{Project}}& \multicolumn{4}{c}{\textbf{\gptturbo}}& \multicolumn{4}{c}{\textbf{\cushman}}& \multicolumn{4}{c}{\textbf{\starcoder}}\\
    \cmidrule(lr){2-5}
        \cmidrule(lr){6-9}
        \cmidrule(lr){10-13}
        & \textbf{Tests} & \textbf{Passing} & \textbf{Stmt Coverage}  & \textbf{Branch Coverage}
        & \textbf{Tests} & \textbf{Passing} & \textbf{Stmt Coverage}  & \textbf{Branch Coverage}
        & \textbf{Tests} & \textbf{Passing} & \textbf{Stmt Coverage}  & \textbf{Branch Coverage}
        \\
    \midrule

glob& 68 & 18 (26.5\%) & \textbf{71.3\%} & \textbf{66.3\%}
        & 76 & 31 (40.1\%) & 61.7\% & 51.0\%
        & 45 & 10 (22.2\%) & 64.8\% & 58.4\%
        \\
        
fs-extra& 471 & 277 (58.8\%) & \textbf{58.8\%} & \textbf{38.9\%}
        & 394 & 254 (64.3\%) & 41.0\% & 23.3\%
        & 443 & 163 (36.7\%) & 43.0\% & 25.5\%
        \\
        
graceful-fs& 345 & 177 (51.4\%) & \textbf{49.3\%} & \textbf{33.3\%}
        & 301 & 135 (44.9\%) & 47.5\% & 30.3\%
        & 309 & 100 (32.4\%) & 44.7\% & 22.7\%
        \\
        
jsonfile& 13 & 6 (48.0\%) & 38.3\% & 29.4\%
        & 15 & 8 (53.3\%) & 46.8\% & 44.1\%
        & 13 & 7 (53.8\%) & \textbf{59.6\%} & \textbf{47.0\%}
        \\
        
bluebird& 370 & 204 (55.2\%) & 68.0\% & 50.0\%
        & 400 & 211 (52.6\%) & \textbf{68.2\%} & \textbf{51.3\%}
        & 395 & 130 (32.8\%) & 55.6\% & 36.0\%
        \\
        
q& 323 & 186 (57.6\%) & \textbf{70.4\%} & \textbf{53.7\%}
        & 356 & 190 (53.4\%) & 66.9\% & 51.2\%
        & 348 & 96 (27.4\%) & 63.0\% & 48.1\%
        \\
        
rsvp& 109 & 70 (64.2\%) & 70.1\% & 55.3\%
        & 115 & 77 (66.5\%) & \textbf{73.3\%} & \textbf{60.5\%}
        & 141 & 45 (31.9\%) & 66.8\% & 53.2\%
        \\
        
memfs& 1037 & 471 (45.4\%) & \textbf{81.1\%} & \textbf{58.9\%}
        & 1058 & 505 (47.7\%) & 78.9\% & 54.9\%
        & 922 & 268 (29.0\%) & 71.9\% & 49.8\%
        \\
        
node-dir& 40 & 19 (48.1\%) & \textbf{64.3\%} & \textbf{50.8\%}
        & 22 & 16 (74.4\%) & 52.2\% & 41.1\%
        & 51 & 17 (33.3\%) & 54.0\% & 42.7\%
        \\
        
zip-a-folder& 11 & 6 (54.5\%) & 84.0\% & 50.0\%
        & 10 & 7 (70.0\%) & \textbf{88.0\%} & \textbf{62.5\%}
        & 11 & 4 (36.4\%) & 56.0\% & 37.5\%
        \\
        
js-sdsl& 409 & 46 (11.3\%) & 33.9\% & 24.3\%
        & 274 & 63 (23.0\%) & \textbf{36.5\%} & \textbf{27.3\%}
        & 235 & 21 (8.9\%) & 26.9\% & 17.9\%
        \\
        
quill-delta& 152 & 33 (21.7\%) & 73.0\% & 64.3\%
        & 187 & 50 (26.5\%) & \textbf{74.0\%} & \textbf{66.6\%}
        & 135 & 7 (5.2\%) & 31.0\% & 21.1\%
        \\
        
complex.js& 209 & 121 (58.0\%) & \textbf{70.2\%} & \textbf{46.5\%}
        & 221 & 125 (56.3\%) & 62.7\% & 46.2\%
        & 178 & 56 (31.5\%) & 53.5\% & 34.9\%
        \\
        
pull-stream& 83 & 34 (41.0\%) & 69.1\% & 52.8\%
        & 76 & 43 (55.9\%) & \textbf{70.8\%} & \textbf{54.7\%}
        & 69 & 10 (14.5\%) & 51.6\% & 32.7\%
        \\
        
countries-and-timezones& 28 & 13 (46.4\%) & \textbf{93.1\%} & 69.1\%
        & 41 & 18 (44.4\%) & \textbf{93.1\%} & \textbf{74.4\%}
        & 33 & 11 (33.8\%) & 88.2\% & 64.9\%
        \\
        
simple-statistics& 353 & 250 (70.9\%) & \textbf{87.8\%} & \textbf{71.3\%}
        & 350 & 213 (60.7\%) & 80.1\% & 63.9\%
        & 352 & 164 (46.6\%) & 69.9\% & 54.5\%
        \\
        
plural& 13 & 8 (61.5\%) & 73.8\% & \textbf{59.1\%}
        & 17 & 8 (47.1\%) & \textbf{75.4\%} & \textbf{59.1\%}
        & 13 & 5 (38.5\%) & 73.8\% & \textbf{59.1\%}
        \\
        
dirty& 70 & 32 (45.3\%) & 74.5\% & 65.4\%
        & 89 & 42 (47.5\%) & \textbf{81.1\%} & \textbf{69.2\%}
        & 57 & 23 (40.4\%) & 72.6\% & 61.5\%
        \\
        
geo-point& 76 & 50 (65.8\%) & \textbf{87.8\%} & \textbf{70.6\%}
        & 87 & 35 (40.2\%) & 61.0\% & \textbf{70.6\%}
        & 62 & 16 (25.8\%) & 46.3\% & \textbf{70.6\%}
        \\
        
uneval& 7 & 2 (28.6\%) & \textbf{68.8\%} & \textbf{58.3\%}
        & 5 & 0 (0.0\%) & 0.0\% & 0.0\%
        & 6 & 0 (0.0\%) & 0.0\% & 0.0\%
        \\
        
image-downloader& 5 & 4 (80.0\%) & 63.6\% & \textbf{50.0\%}
        & 5 & 2 (40.0\%) & \textbf{75.8\%} & \textbf{50.0\%}
        & 5 & 2 (40.0\%) & 63.6\% & \textbf{50.0\%}
        \\
        
crawler-url-parser& 14 & 2 (14.3\%) & \textbf{51.4\%} & \textbf{35.0\%}
        & 17 & 2 (11.8\%) & 49.5\% & 31.3\%
        & 14 & 1 (7.1\%) & 48.6\% & 32.5\%
        \\
        
gitlab-js& 141 & 14 (9.9\%) & 51.7\% & 16.5\%
        & 116 & 35 (29.7\%) & \textbf{61.8\%} & \textbf{31.8\%}
        & 117 & 1 (0.9\%) & 28.4\% & 0.6\%
        \\
        
core& 85 & 13 (15.3\%) & \textbf{78.3\%} & \textbf{50.0\%}
        & 102 & 21 (20.7\%) & 72.7\% & 47.7\%
        & 61 & 5 (8.2\%) & 16.1\% & 0.0\%
        \\
        
omnitool& 1033 & 330 (31.9\%) & \textbf{74.2\%} & \textbf{55.2\%}
        & 1029 & 321 (31.1\%) & 70.1\% & 54.2\%
        & 812 & 194 (23.9\%) & 40.0\% & 18.1\%
        \\
        \midrule
\textbf{Median} & 
    & \textbf{48.0\%} 
    & \textbf{70.2\%}
    & \textbf{52.8\%}& 
    & 47.1\% 
    & 68.2\%
    & 51.2\%& 
    & 31.5\% 
    & 54.0\%
    & 37.5\%\\
\bottomrule
\end{tabular}
}
\vspace{-0.5cm}
\end{table*}

Figure~\ref{fig:cumulative_sim} shows the cumulative percentage of generated tests cases for each project where the maximum similarity is less than the value on the x-axis.
We also show this cumulative percentage for \textit{all} generated test cases across all projects.
We find that \gptturbosimLEQthirty of \testpilot's generated test cases have less than $\leq 0.3\%$ maximum similarity to an existing test, \gptturbosimLEQforty have $\leq 0.4$ similarity, \gptturbosimLEQfifty have $\leq 0.5$, \gptturbosimLEQsixty have $\leq 0.6$ while \gptturbosimLEQseventy of the generated tests cases have $\leq 0.7$.
This means that \testpilot \textbf{never} generates exact copies of existing tests.
In contrast, while 90\% of Lemieux et al.~\cite{LemieuxICSE2023}'s generated Python tests have $\leq 0.4$ similarity, 2\% of their test cases are exact copies.
That said, given the differences between testing frameworks in Python and JavaScript (e.g., Mocha requires more boilerplate code than pytest), similarity numbers cannot be directly compared between the two languages.

To further illustrate the resulting similarity numbers, Figure~\ref{fig:highSimilarity} shows an example of a test case from \texttt{bluebird} with 0.62 similarity to an existing test case.
While the edit distance here is low, resulting in the high similarity, we can see that the tests have semantic differences.
For example, the generated test simply checks that the thrown exception is a type error, while the existing test checks for certain values in the trace.
Thus, the \gptturbosimGTfifty of test cases we generate with $> 0.5$ similarity do not pose a concern that \testpilot is generating memorized test cases.
Finally, we would expect the generated tests for GitLab-hosted projects to have a lower similarity to existing tests since, as far as we know, the training set for OpenAI's models only includes projects from GitHub, so the model is less likely to have seen the existing tests during training.
Our results do indeed show that three out of the five projects have a maximum similarity of $\leq 0.4$, with the remaining two having maximum similarity of 0.5.
This gives us confidence that the similarity metric we use provides meaningful results.

\subsection{\ref{rq:llm-comparison}: Effect of Different LLMs}

Table~\ref{tab:llm-comparison} shows the number of generated tests, percent of generated tests that pass, as well as statement and branch coverage of \testpilot's generated tests when using three different LLMs.
While the individual coverage per package varies, we can see that the coverage of tests generated by the \cushman model is comparable to those generated by \gptturbo, with the latter having a slightly higher median statement and branch coverage across the packages.
A Wilcoxon matched-pairs signed-rank test shows no statistically significant differences between \gptturbo and \cushman for either type of coverage.
On the other hand, we do find a statistically significant difference between \starcoder and each of the OpenAI models (p-value $< 0.05$) for both types of coverage.
As shown in Table~\ref{tab:llm-comparison}, \starcoder achieves lower median statement (\starcodermedianstmtCoverage) and branch coverage (\starcodermedianbranchCoverage) than both other models.
Cliff's delta~\cite{cliff1993dominance} shows a large and medium effect size for statement and branch coverage, respectively, between \gptturbo and \starcoder and a medium and small effect size for statement and branch coverage, respectively between \cushman and \starcoder.

However, we note that \starcoder's median statement coverage and branch coverage are both higher than Nessie (statement: \starcodermedianstmtCoverage vs. \starcodermedianNessiestmtCoverage and branch: \starcodermedianbranchCoverage vs \starcodermedianNessiebranchCoverage).
While this higher coverage was not statistically significant, the results show that even LLMs trained with potentially smaller datasets and/or a different training process than OpenAI's models are on par (or even sometimes higher) than state-of-the-art traditional test-generation techniques, such as Nessie~\cite{DBLP:conf/icse/ArtecaHPT22}.
Furthermore, in~\ref{rq:nessie-comparison}, we showed that using \gptturbo with \testpilot resulted in higher coverage test suites, with statistically significant differences to Nessie.
Overall, these results emphasize the promise of LLM-based test generation techniques in generating high coverage test suites.

Finally, we note that the median time for \testpilot to generate tests for a given function using \gptturbo is \gptturbomedianAvgTimePerMethod, and the median time to generate a complete test suite for a given package is \gptturbomedianTotalTime.%
\footnote{These timings were measured on a standard GitHub Actions Linux VM with a 2-core CPU, 7GB of RAM, and 14GB of SSD disk space.}
The bulk of this time is spent querying the model, so the choice of LLM makes a big difference.
For example, the median time for \testpilot to generate tests for a given function using \starcoder and \cushman is \starcodermedianAvgTimePerMethod and \cushmanmedianAvgTimePerMethod, respectively, and \starcodermedianTotalTime and \cushmanmedianTotalTime, respectively, for a complete test suite.
All these performance numbers suggest that it is feasible to use \testpilot either in an online setting (e.g., in an IDE) to generate tests for individual functions, or in an offline setting (e.g., during code review) to generate complete test suites for an API.

\section{Threats to Validity}

\paragraph*{\textbf{Internal Validity}}
The extraction of example snippets from documentation relies on textually matching a function's name.
Given two functions with the same name but different access paths, we cannot disambiguate which function is being used in the example snippet.
Accordingly, we match this snippet to both functions.
While this may lead to inaccuracies, there is unfortunately no precise alternative for this matching.
Any heuristics may cause us to miss snippets altogether, which may be worse since example snippets help with increasing the percentage of passing tests as shown in Figure~\ref{fig:refiner-boxplot}.
The overall high coverage and percentage of passing tests suggest that our matching technique is not a limiting factor in practice.

\paragraph*{\textbf{Construct Validity}} 
We use the concept of non-trivial assertions as a proxy for oracle quality in the generated tests.
When determining non-trivial assertions, we search for \textit{any} usage of the package under test in the backwards slice of the assertion. Such usage may be different from the intended function under test.
However, given the dynamic nature of JavaScript, precisely determining the usage of a given function, as extracted by the API explorer, and its occurrence in the backwards slice is difficult.
While our approach does not allow us to precisely determine non-trivial coverage for a given function, this does not affect the non-trivial coverage we report for each package's complete API.
Note that when calculating non-trivial coverage, we measure the full coverage of tests that contain \textit{at least} one non-trivial assertion.
There may be other calls in those non-trivial tests that contribute to coverage but do not contribute to the assertion.
Measuring assertion/checked coverage as defined by Schuler and Zeller~\cite{SchulerCheckedCoverage} is a possible alternative, 
but this is practically difficult to implement precisely for JavaScript.

Our definition of non-trivial assertions is simple, setting a low bar for non-triviality. Any assertion classified as trivial by our criterion is, indeed, not meaningful,
but the converse is not necessarily true. Accordingly, our measure of non-trivial coverage is a lower bound on the true non-trivial coverage.

While the examples we show in the paper suggest that \testpilot's generated tests use variable names that are similar to those chosen by programmers, we do not formally assess the readability of these tests. 
In the future, it would be interesting to conduct user studies to assess the readability of tests generated by different techniques.

\paragraph*{\textbf{External Validity}} 
Despite our evaluation scale significantly exceeding evaluations of previous test generation approaches~\cite{DBLP:conf/icse/ArtecaHPT22, bareiss22}, our results are still based on \gptturbonumPackages npm packages and may not generalize to other JavaScript code bases.
In particular, \testpilot's performance may not generalize to proprietary code that was never seen in the LLM's training set.
We try to mitigate this effect in several ways: (1) we evaluate on less popular packages that are likely to have had less impact on the model's training, (2) we evaluate on 5 GitLab repositories that have not been included in the models' training, and (3) we measure the similarity of the generated tests to the existing tests.
Our results show that \testpilot performs well for both popular and unpopular packages and that \gptturbosimLEQfifty of the test cases have $\leq$ 50\% similarity with existing tests, with no exact copies. Overall, this reassures us that \testpilot is not producing ``memorized'' code.

Finally, we note that while our technique is conceptually language-agnostic, our current implementation of \testpilot targets JavaScript, and thus
we cannot generalize our results to other languages.

\section{Related work}\label{sec:related}

\testpilot provides an alternative to (and potentially complements) traditional techniques for automated test generation, including
  feedback-directed random test generation \cite{Csallner2004,Pacheco:2007,Pacheco:2008:FEN:1390630.1390643,DBLP:journals/pacmpl/SelakovicPKT18,DBLP:conf/icse/ArtecaHPT22},
  search-based and evolutionary techniques \cite{DBLP:conf/qsic/FraserA11,DBLP:conf/sigsoft/FraserA11,DBLP:conf/icst/McMinn11,PanichellaMOSA2018}, and
  dynamic symbolic execution \cite{DBLP:conf/pldi/GodefroidKS05,DBLP:conf/sigsoft/SenMA05,DBLP:conf/ccs/CadarGPDE06,DBLP:conf/kbse/TillmannHX14}.
This section reviews neural techniques for test generation, and previous test generation techniques for JavaScript.

\subsection{Neural Techniques}

Neural techniques are rapidly being adopted for solving various Software Engineering problems, with promising
results in several domains including
  code completion~\cite{DBLP:conf/ijcai/LiWLK18,DBLP:conf/sigsoft/SvyatkovskiyDFS20,DBLP:conf/icse/KarampatsisBRSJ20,DBLP:conf/icse/KimZT021,CopilotSite}, 
  program repair~\cite{DBLP:conf/aaai/GuptaPKS17,DBLP:conf/iclr/HellendoornSSMB20,DBLP:conf/nips/AllamanisJB21}, and 
  bug-finding~\cite{DBLP:journals/pacmpl/PradelS18,DBLP:conf/iclr/AllamanisBK18}. 
Pradel and Chandra~\cite{DBLP:journals/cacm/PradelC22} survey the current state of the art in this emerging research area.
 We are aware of several recent research efforts in which LLMs are used for test generation~\cite{tappy,codet,bareiss22,tufano2021unit,LemieuxICSE2023,atlas,MastrapaoloICSE2021}.
There are two main differences between our work and these efforts: (i) the goal and types of tests generated and (ii) the need for some form of fine-tuning or additional data.
We discuss the details below.

\paragraph*{\textbf{Differing goals}} \textsc{TiCoder}~\cite{tappy} and \textsc{CodeT}~\cite{codet} use Codex to generate implementations
and test cases from problem descriptions expressed in natural language. \textsc{TiCoder} relies on a test-driven user-intent
formalization (TDUIF) loop in which the user and model interact to generate both an implementation matching 
the user's intent and a set of test cases to validate its correctness. \textsc{CodeT}, on the other
hand, generates both a set of candidate implementations and some test cases based on the same prompt, runs the generated tests on the candidate implementations, and chooses the best solution based on the test results.
Unlike \testpilot, neither of these efforts solves the problem of automatically generating unit tests for \textit{existing code}.

Given the characteristics of LLMs in generating natural looking code, there have been several efforts exploring the use of LLMs to help~\cite{LemieuxICSE2023} or complement~\cite{atlas} traditional test generation techniques.
Most recently, Lemieux et al.~\cite{LemieuxICSE2023} explore using tests generated by Codex as a way to unblock the search process of test generation using search-based techniques~\cite{PanichellaMOSA2018}, which often fails when the initial randomly generated test has meaningless input that cannot be mutated effectively. 
Their results show that, on most of their target 27 Python projects, their proposed technique, \textsc{CodaMosa}, outperforms the baseline search-based technique, Pynguin's implementation of \textsc{Mosa}~\cite{PanichellaMOSA2018}, as well as using only Codex.
However, their Codex prompt includes only the function implementation and an instruction to generate tests. Since their main goal is to explore whether a test generated by Codex can improve the search process, they do not systematically explore the effect of different prompt components. In fact, they conjecture that further prompt engineering might improve results, motivating the need for our work which systematically explores different prompt components. Additionally, their generated tests are in the MOSA format~\cite{PanichellaMOSA2018}, which the authors acknowledge could lose readability, and do not contain assertions. Most of our tests contain assertions, and we further study the quality of assertions we generate as well as reasons for test failures.

Similarly, given that it is often difficult for traditional test generation techniques to generate (useful) assertions~\cite{DBLP:conf/icsm/PanichellaPFSH20,  DBLP:conf/icse/PalombaNPOL16}, \textsc{Atlas}~\cite{atlas} uses LLMs to generate an assert statement for a given (assertion-less) Java test.
They position their technique as a complement to traditional techniques~\cite{DBLP:conf/sigsoft/FraserA11,Pacheco:2007}.
With the same goal, Mastrapaolo et al.~\cite{MastrapaoloICSE2021,MastrapaoloTSE2022} and Tufano et al.~\cite{TufanoAST22Assert} perform follow up work using transfer learning, while Yu et al.~\cite{yu22} use information retrieval techniques to further improve the assert statements generated by Atlas.
TOGA~\cite{togaICSE22} uses similar techniques but additionally incorporates an exceptional oracle classifier to decide if a given method requires an assertion to test exceptional behavior.
It then bases the generation of the assertion on a pre-defined oracle taxonomy created by manually analyzing existing Java tests and using a neural-based ranking mechanism to rank candidates with oracles higher.
In contrast with these efforts, our goal is to generate a \textit{complete} test method without giving the model any content of the test method (aside from boilerplate code required by Mocha), which means that the model needs to generate \textit{both} any test setup code (e.g., initializing objects and populating them) as well as the assertion.
While TOGA can be integrated with EvoSuite~\cite{DBLP:conf/qsic/FraserA11} to create an end-to-end test-generation tool, recent work~\cite{LiuISSTA23Eval} points out several shortcomings of the evaluation methods, casting doubt on the validity of the reported results.

\paragraph*{\textbf{Differing Input/Training}} Barei{\ss} et al.~\cite{bareiss22} evaluate the performance of Codex on three code-generation tasks, including test
generation. Like us, they rely on embedding contextual information into the prompt to guide the LLM, 
though the specific data they embed is different: while \toolname\  only includes the signature, definition, 
documentation, and usage examples in the prompt,  Barei{\ss} et al. pursue a few-shot learning approach where, in
addition to the definition of a function under test, they include an example of a different function from the same code base and its associated 
test to give the model a hint as to what it is expected to do, as well as a list of related helper function signatures that could be useful for test generation. For a limited list of 18 Java methods, they show that this approach yields 
slightly better coverage than Randoop~\cite{Pacheco:2007,Pacheco:2008:FEN:1390630.1390643}, a popular technique for feedback-directed random test generation.
This is a promising result, but finding suitable example tests to use in few-shot learning can be difficult, especially since their evaluation shows that
good coverage crucially depends on the examples being closely related to the function under test.%
 
Tufano et al. \cite{tufano2021unit} present AthenaTest, an approach for automated test generation based on a BART
transformer model \cite{lewis2019bart}. For a given test case, they rely on heuristics to identify the ``focal'' class and method 
under test. These mapped test cases are then used to fine-tune the model for the task of producing unit tests by 
representing this task as a translation task that maps a focal method (along with the focal class, constructors, and other public
methods and fields in that class) to a test case. In experiments on 5 projects from Defects4J \cite{DBLP:conf/issta/JustJE14},
AthenaTest generated 158K test cases, achieving similar test coverage as EvoSuite \cite{DBLP:conf/qsic/FraserA11}, a popular search-based test generation tool,
and covering 43\% of all focal methods.  
A significant difference between their work and ours is that their approach requires training the model on a large set
of test cases whereas \testpilot uses an off-the-shelf LLM. In fact, in addition to the goal differences with ATLAS~\cite{atlas} and Mastrapaolo et al.'s~\cite{MastrapaoloICSE2021, MastrapaoloTSE2022} work above, both these efforts also require a data set of test methods (with assertions) and their corresponding focal methods, whether to use in the main training~\cite{atlas} or in fine tuning during transfer learning~\cite{MastrapaoloICSE2021,MastrapaoloTSE2022,TufanoAST22Assert}.

Unfortunately, the above differences in goals or in the required data for model training make it meaningless or impossible to do a direct experimental comparison with \testpilot. 
Additionally, none of these efforts support JavaScript or provide JavaScript data sets that can be used for comparison.
In fact, one of our main motivations for exploring prompt engineering for an off-the-shelf LLM is to avoid the need to collect test examples for few-shot learning~\cite{bareiss22} or test method/focal method pairs required for training~\cite{atlas} or additional fine tuning~\cite{MastrapaoloICSE2021,MastrapaoloTSE2022,TufanoAST22Assert}.

\paragraph*{\textbf{Other techniques}}

Stallenberg et al. \cite{DBLP:conf/ssbse/StallenbergOP22} present a test generation technique for JavaScript based on unsupervised type inference
consisting of three phases. First, a static analysis is performed to deduce
relationships between program elements such as variables and expressions. Then, a probabilistic type inference is applied to these 
relationships to construct a model.
Finally, they show how search-based techniques can take advantage of the information contained in such models by proposing
 two strategies for consulting these models in the main loop of DynaMOSA \cite{PanichellaMOSA2018}.
 
Recently, El Haji \cite{ElHajiThesis} presented an empirical study that explores the effectiveness of GitHub Copilot at generating tests. In this study, tests 
are selected from existing test suites associated with 7 open-source Python projects. After removing the body of each test function, Copilot is asked to 
complete the implementation so that the resulting test can be executed and compared against the original test. Two variations of this approach are 
explored, viz., ``with context’’ where the other tests in the suite are preserved and ``without context’’ where other tests are removed. El Haji also explores 
the impact of (manually) adding comments that include descriptions of intended behavior and usage examples. The results from the study show that 45.28\% of 
generated test are passing in the ``with context’’ scenario (the rest are failing, syntactically invalid, or empty) vs only 7.55\% passing generated tests 
in the ``without context’’ scenario, and that the addition of usage examples and comments is generally helpful. 
There are several significant differences between our approach and El Haji’s work:
  we explore a fully automated technique without any manual steps,
  we report on a significantly more extensive empirical evaluation,
  we present an adaptive technique in which prompts are refined in response to the execution behavior of previously executed tests,
  we target a different programming language (JavaScript instead of Python), and 
  TestPilot interacts directly with an LLM rather than relying on Copilot, an LLM-based programming assistant. 
 
\subsection{Test Generation Techniques for JavaScript}

\testpilot's mechanism for refining prompts based on execution feedback was inspired by the mechanism employed by 
feedback-directed random test generation techniques \cite{Csallner2004,Pacheco:2007,Pacheco:2008:FEN:1390630.1390643,DBLP:journals/pacmpl/SelakovicPKT18,DBLP:conf/icse/ArtecaHPT22}, 
where new tests are generated by extending previously generated passing tests. As reported in Section~\ref{sec:NessieComparison}, \testpilot achieves significantly higher statement coverage
and branch coverage than Nessie \cite{DBLP:conf/icse/ArtecaHPT22}, which represents the state-of-the-art in feedback-directed random test generation for JavaScript. 

Several previous projects have considered test generation for JavaScript (see \cite{DBLP:journals/csur/AndreasenGMPSSS17} for a survey).
Saxena et al. \cite{DBLP:conf/sp/SaxenaAHMMS10} present Kudzu, a tool that aims to find injection vulnerabilities in 
client-side JavaScript applications by exploring an application's input space. They differentiate an application's 
input space into an \textit{event space}, which concerns the order in which event handlers execute (e.g., as a result of 
buttons being clicked), and a \textit{value space} which concerns the choice of values passed to functions or entered 
into text fields. Kudzu uses dynamic symbolic execution to explore the value space systematically, but it relies on a random 
exploration strategy to explore the event space.   Artemis \cite{DBLP:conf/icse/ArtziDJMT11} is a framework for automated test 
generation that iteratively generates tests for client-side JavaScript applications consisting of sequences of events, using a 
heuristics-based strategy that considers the locations read and written by each event handler to focus on the generation of tests 
involving event handlers that interact with each other. Li et al. \cite{DBLP:conf/sigsoft/LiAG14} extends Artemis with dynamic 
symbolic execution to improve  its ability to explore the value space, and Tanida et al. \cite{TanidaEtAl:2015} further improve 
on this work by augmenting generated test inputs with user-supplied invariants.  
Fard et al. \cite{DBLP:conf/kbse/Fard0W15} present ConFix, a tool that uses a combination of dynamic
analysis and symbolic execution to automatically generate instances of the Document Object Model (DOM) that can serve 
as test fixtures in unit tests for client-side JavaScript code. 
Marchetto and Tonella \cite{DBLP:journals/ese/MarchettoT11} present a search-based test generation technique that constructs
tests consisting of sequences of events that relies on the automatic extraction of a finite state machine that
represents that application's state.
None of these tools generate tests that contain assertions.   

Several test generation tools for JavaScript are capable of generating tests containing assertions.
JSART \cite{DBLP:conf/icwe/MirshokraieM12} is a tool that generates regression tests that contain
assertions reflecting likely invariants that are generated using a variation of the Daikon dynamic 
invariant generator \cite{DaikonSite}. Since Daikon generates assertions that are \textit{likely} to hold,
an additional step is needed in which invalid assertions are removed from the generated tests.
Mirshokraie et al. \cite{DBLP:conf/kbse/MirshokraieMP13,DBLP:conf/icst/Mirshokraie0P15} present an approach 
in which tests are generated for client-side JavaScript applications consisting of sequences of events.
Then, in an additional step, function-level unit tests are derived  by instrumenting program execution to monitor the state of parameters, 
global variables, and the DOM upon entry and exit to functions to obtain values with which functions are to be invoked. 
Assertions are added automatically to the generated tests by:
  (i) mutating the DOM and the code of the application under test, 
  (ii) executing generated tests to determine how application state is impacted by mutations, and 
  (iii) adding assertions to the tests that reflect the behavior prior to the mutation.     
Testilizer \cite{DBLP:conf/kbse/FardMM14} is a test generation tool that aims to enhance an existing human-written
test suite. To this end, Testilizer instruments code to observe how existing tests access the DOM, 
and executes them to obtain a State-Flow Graph in which the nodes reflect dynamic DOM states and edges reflect the 
event-driven transitions between these states. Alternative paths are explored by exploring previously unexplored
events in each state. Testilizer adds assertions to the generated tests that are either copied verbatim from 
existing tests, by adapting the structure of an existing assertion to a newly explored state, or by inferring
a similar assertion using machine learning techniques.

These techniques share the limitation that they require the entire application under the test to be executable,
limiting their applicability.  Moreover, several of the techniques discussed above require re-execution of tests
(to infer assertions using mutation testing \cite{DBLP:conf/kbse/MirshokraieMP13,DBLP:conf/icst/Mirshokraie0P15}, or 
to filter out assertions that are invalid \cite{DBLP:conf/icwe/MirshokraieM12}), which adds to their cost.
By contrast, \testpilot only requires the functions of API functions under test to be available and executable,
and it executes each test that it generates only once.

\vspace{-0.2cm}
\section{Conclusions and Future Work}
We have presented \testpilot, an approach for adaptive unit-test generation
using a large language model. Unlike previous work in this area, \testpilot{}
requires neither fine tuning nor a parallel corpus of functions and tests.
Instead, we embed contextual information about the function under test into the
prompt, specifically its signature, its attached documentation comment (if any),
any usage examples from the project documentation, and the source code of the
function. Furthermore, if a generated test fails, we adaptively create a new
prompt embedding this test and the failure message to guide the model towards
fixing the problematic test.
We have implemented our approach for JavaScript on top of off-the-shelf LLMs,
and shown that it achieves state-of-the art statement coverage
on \gptturbonumPackages npm packages. Further evaluation shows that the majority of the generated tests
contain non-trivial assertions, and that all parts of the information included
in the prompt contributes to the quality of the generated tests.
Experiments with three LLMs (\gptturbo, \cushman, and \starcoder) demonstrate that
LLM-based test generation already outperforms state-of-the-art previous test generation methods such as Nessie on key metrics. 
We conjecture that the use of more advanced LLMs will further improve results, though we are reluctant to speculate by how much.

In future work, we plan to further investigate the quality of the tests
generated by \testpilot{}. While in this paper we have focused on passing tests and
excluded failing tests from consideration entirely, we have seen examples of
failing tests that are ``almost right'' and might be interesting to a developer
as a starting point for further refinement. However, doing this depends on
having a good strategy for telling apart useful failing tests from useless ones.
Our experiments have demonstrated that timeout errors, assertion errors, and correctness 
errors are key factors that cause tests to fail. In future work, we plan to apply
static and dynamic program analysis to failing tests in order to determine why timeout 
errors and assertion errors occur and how failing tests could be modified to make them pass. 
 
Further research is needed to determine what factors prevent the generation of non-trivial assertions. 
Anecdotally, we have observed that the availability of usage examples is generally helpful. 
We envision that the number of useful assertions could be improved by obtaining usage examples 
in other ways, e.g., by interacting with a user, or by extracting usage examples from clients of 
the application under test.

Another fruitful area of experimentation could be varying the
sampling temperature of the LLM. In this work, we always sample at
temperature zero, which has the advantage of providing stable results, but also
means that the model is less likely to offer lower-probability completions that
might result in more interesting tests.

Another area of future work is the development of hybrid techniques that
combine existing feedback-directed test generation techniques with an LLM-based
technique such as \testpilot. For example, one could use an LLM-based
technique to generate an initial set of tests and use the tests that it generates 
as a starting point for extension by a feedback-directed technique such as Nessie,
thus enabling it to uncover edges cases that would be difficult to uncover
using only random values.

In principle, our approach can be adapted to any programming language. Practically
speaking, this would involve adapting prompts to use the syntax of the language
under consideration, and to use a testing framework for that language. In addition,
the mining of documentation and usage examples would need to be adapted to 
match the documentation format used for that language. The LLMs that 
we used did not language-specific training and could be used to generate tests for
other languages, though the effectiveness of the approach will depend on the
amount of code written in that language that was included in the LLM's training
set. One specific question that would be interesting to explore is how well
an approach like \testpilot{} would perform on a statically-typed language.

\section*{Acknowledgment}
The research reported on in this paper was conducted while S. Nadi and F. Tip were sabbatical visitors and A. Eghbali an intern at GitHub. The authors are grateful to the GitHub Next team for many insightful and helpful discussions about TestPilot. F. Tip was supported in part by National Science Foundation grants CCF-1907727 and CCF-2307742. S. Nadi's research is supported by the Canada Research Chairs Program and the Natural Sciences and Engineering Research Council of Canada (NSERC), RGPIN-2017-04289.

\bibliographystyle{IEEEtran}
\bibliography{references}
\end{document}